\documentclass{elsarticle}

\usepackage[reviewcopy]{adndt}%{adndt} at NSCL, {../adndt} at Gross
\usepackage{longtable}

\usepackage{amsmath}

\usepackage{amssymb}

\biboptions{square,sort&compress}
\bibpunct[]{[}{]}{,}{n}{}{;}
\begin{document}

\begin{frontmatter}

\journal{Atomic Data and Nuclear Data Tables}

%% Author, fill in article title here

\title{Discovery of Calcium, Indium, Tin, and Platinum Isotopes}

%% Fill in author list here
 \author{S. Amos}
 \author{J. L. Gross}
 \author{M. Thoennessen\corref{cor1}}\ead{thoennessen@nscl.msu.edu}

 \cortext[cor1]{Corresponding author.}

 \address{National Superconducting Cyclotron Laboratory and \\ Department of Physics and Astronomy, Michigan State University, \\ East Lansing, MI 48824, USA}

\begin{abstract}
Currently, twenty-four calcium, thirty-eight indium, thirty-eight tin and thirty-nine platinum isotopes have been observed and the discovery of these isotopes is discussed here. For each isotope a brief synopsis of the first refereed publication, including the production and identification method, is presented.
\end{abstract}

\end{frontmatter}

%%% Keywords and subject classification are not used in ADNDT
%%%\begin{keywords}
%%%Insert list of keywords here.
%%%\end{keywords}

%%%\begin{subject}[Insert header for classifications]
%%%Use only if your journal has a subject classification requirement
%%%\end{subject}

%%% The table of contents should start a new page. This command will
%%% automatically pull all the unstarred \section, \subsection and
%%% \subsubsection titles into the Contents. Starred versions need to be
%%% done manually using
%%%      \addcontentsline{toc}{[[sub]sub]section}{Section title}
%%% at the correct place. Examples are given below.

%%% The lists of data figures and data tables are created automatically
%%% by the \listofDfigures and \listofDtables commands.

\newpage
\tableofcontents
%%\listofDfigures
\listofDtables

\vskip5pc

\section{Introduction}\label{s:intro}

The discovery of calcium, indium, tin, and platinum isotopes is discussed as part of the series summarizing the discovery of isotopes, beginning with the cerium isotopes in 2009 \cite{2009Gin01}. Guidelines for assigning credit for discovery are (1) clear identification, either through decay-curves and relationships to other known isotopes, particle or $\gamma$-ray spectra, or unique mass and Z-identification, and (2) publication of the discovery in a refereed journal. If the first observation was not confirmed or found erroneous in the subsequent literature, the credit was given to the correct measurement. These cases are specifically mentioned and discussed. Thus the assignment for the more recent papers is subject to confirmation. The authors and year of the first publication, the laboratory where the isotopes were produced as well as the production and identification methods are discussed. When appropriate, references to conference proceedings, internal reports, and theses are included. When a discovery includes a half-life measurement the measured value is compared to the currently adopted value taken from the NUBASE evaluation \cite{2003Aud01} which is based on the ENSDF database \cite{2008ENS01}. In cases where the reported half-life differed significantly from the adapted half-life (up to approximately a factor of two), we searched the subsequent literature for indications that the measurement was erroneous. If that was not the case we credited the authors with the discovery in spite of the inaccurate half-life.

\section{Discovery of $^{35-58}$Ca}

Twenty four calcium isotopes from A = $35-58$ have been discovered so far; these include 6 stable, 6 proton-rich and 12 neutron-rich isotopes.  According to the HFB-14 model \cite{2007Gor01}, $^{63}$Ca should be the last odd-even particle stable neutron-rich nucleus while the even-even particle stable neutron-rich nuclei should continue at least through $^{70}$Ca. At the proton dripline two more isotopes could be observed ($^{33}$Ca and $^{34}$Ca). About 11 isotopes have yet to be discovered corresponding to 30\% of all possible calcium isotopes.

\begin{figure}
	\centering
	\includegraphics[scale=.5]{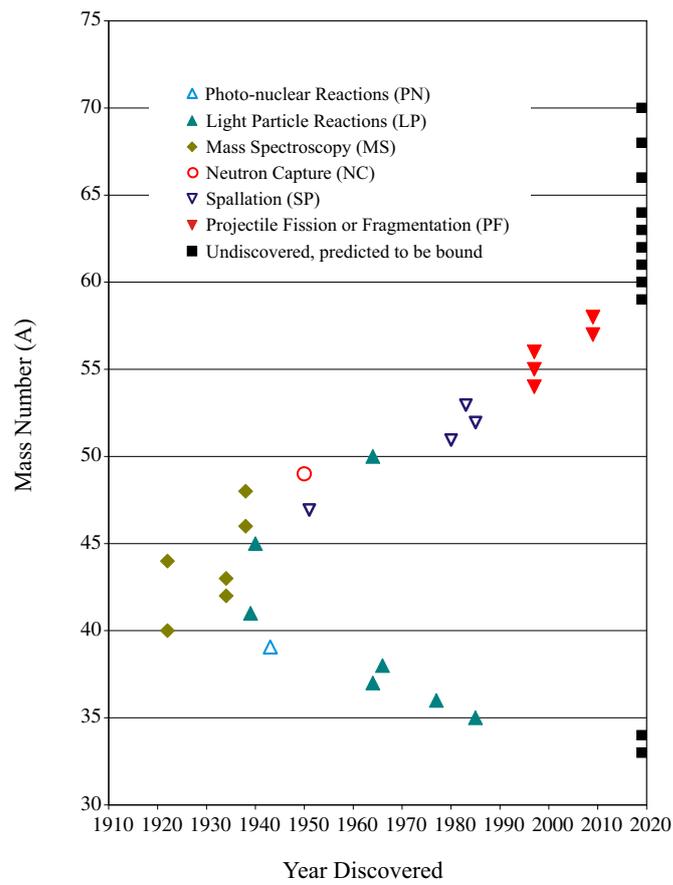}
	\caption{Calcium isotopes as a function of time when they were discovered. The different production methods are indicated. The solid black squares on the right hand side of the plot are isotopes predicted to be bound by the HFB-14 model.}
\label{f:year-ca}
\end{figure}

Figure \ref{f:year-ca} summarizes the year of first discovery for all calcium isotopes identified by the method of discovery. The range of isotopes predicted to exist is indicated on the right side of the figure. The radioactive calcium isotopes were produced using photo-nuclear  reactions (PN), neutron capture reactions (NC), light-particle reactions (LP), spallation (SP), and projectile fragmentation or fission (PF). The stable isotopes were identified using mass spectroscopy (MS). Light particles also include neutrons produced by accelerators. In the following, the discovery of each calcium isotope is discussed in detail and a summary is presented in Table 1.

\subsection*{$^{35}$Ca}\vspace{0.0cm}

$^{35}$Ca was discovered by \"Ayst\"o et al. in 1985, and published in ``Observation of the First T$_{z}$=$-\frac{5}{2}$ Nuclide, $^{35}$Ca, via Its $\beta$-Delayed Two-Proton Emission'' \cite{1985Ays01}.  A beam of 135-MeV $^{3}$He from the Berkeley 88-inch cyclotron bombarded a 2-mg/cm$^2$ natural-calcium target. The $\beta$-delayed two-proton sum spectra were measured and assigned to $^{35}$Ca.  ``The assignment of the observed groups to $^{35}$Ca is based on excellent agreement with the predicted decay energy for the higher sum peak populating the $^{33}$Cl ground state and with the known energy difference for decays to the ground (\textit{G}) and the first excited (\textit{X}) states at 811 keV in $^{33}$Cl. Further, the half-life is consistent with the predictions for $^{35}$Ca and no other new beta-delayed two-proton emitters (e.g., $^{27}$S), if produced, are expected to have these two-proton sum energies.'' The measured half-life of 50(30)~ms agrees with the currently adopted value of 25.7(2)~ms.

\subsection{$^{36}$Ca}\vspace{0.0cm}

Tribble et al. first observed $^{36}$Ca in 1976. They reported their findings in ``Mass of $^{36}$Ca'' \cite{1977Tri01}. A 121.4-MeV $\alpha$ beam from the Texas A\&M University 88-inch cyclotron bombarded a 3-mg/cm$^2$ natural calcium target. The presence of $^{36}$Ca was inferred from the presence of $^8$He detected by an Enge split-pole magnetic spectrograph. ``The centroid uncertainty, assuming background contribution, is 30 keV. Combining this with  the uncertainties associated with (1) beam energy (10 keV), (2) scattering angle (5 keV), (3) focal plane calibration (15 keV), target thickness (20 keV) along with the $^8$He mass excess of 31.601$\pm$0.013 MeV, we find the reaction Q value to be -57.58$\pm$0.04 MeV, and the mass of $^{36}$Ca to be -6.44$\pm$0.04 MeV.''

\subsection{$^{37}$Ca}\vspace{0.0cm}

The discovery of $^{37}$Ca was simultaneously reported in 1964 by Hardy and Verrall in \textit{Calcium-37} \cite{1964Har01} and Reeder et al. in ``New Delayed-Proton Emitters: Ti$^{41}$, Ca$^{37}$, and Ar$^{33}$'' \cite{1964Ree01}. Hardy and Verrall bombarded a calcium target with an 85 MeV proton beam from the McGill synchrocyclotron. The delayed proton spectrum was measured with a surface barrier silicon detector to identify the presence of $^{37}$Ca. ``The threshold for production from stable calcium (97\% $^{40}$Ca) was found to be 7 MeV higher than  that from potassium (93\% $^{39}$K), and was approximately 47 MeV. These results are compatible only with the reactions $^{40}$Ca(\textit{p},\textit{d}2\textit{n})$^{37}$Ca and $^{39}$K(\textit{p},3\textit{n})$^{37}$Ca, whose calculated laboratory energy thresholds are 44.6 and 38.5 MeV. This establishes the activity as following the decay of $^{37}$Ca.'' Reeder et al. used the 60-in. cyclotron at Brookhaven to bombard gaseous $^{36}$Ar with $^3$He at a maximum energy of 31.8 MeV. Proton spectra were measured by two surface barrier detectors. ``The excitation function observed for Ca$^{37}$ has a threshold at 20$\pm$2 MeV which is consistent with the predicted threshold of 19.4 MeV for the (He$^3$,2\textit{n}) reaction.'' The papers were submitted on the same day and published in the same issue of Physical Review Letters.

\subsection{$^{38}$Ca}\vspace{0.0cm}

The discovery of $^{38}$Ca was reported in 1966 by Hardy et al. in ``Energy Levels of$^{38}$Ca From the Reaction $^{40}$Ca(p,t)$^{38}$Ca'' \cite{1966Har01}. A 39.8 MeV beam from the Rutherford Laboratory Proton Linear Accelerator bombarded natural calcium targets. A semi-conductor counter telescope was used to detect the emitted particles. The Q value for the $^{40}$Ca(p,t)$^{38}$Ca reaction was measured and a mass excess was calculated for $^{38}$Ca. ``The value obtained for the $^{40}$Ca(p,t)$^{38}$Ca Q value is -20.459$\pm$0.025 MeV.'' In 1957 a half-life measurement for $^{38}$Ca of 0.66~s produced in the $^{40}$Ca($\gamma$,2n) reaction was based on the observation of a 3.5 MeV $\gamma$ \cite{1957Cli01} which could not be confirmed \cite{1968Kav01}. Another experiment using the same reaction relied on the 1957 measurement and did not identify $^{38}$Ca independently \cite{1966And01}.

\subsection{$^{39}$Ca}\vspace{0.0cm}

$^{39}$Ca was first observed in 1943 by Huber et al.: ``Der Kernphotoeffekt mit der Lithium-Gammastrahlung: I. Die leichten Elemente bis zum Calcium'' \cite{1943Hub01}. $^{39}$Ca was populated in a radiative capture reaction with 17 MeV $\gamma$-rays. 500 keV protons bombarded lithium to produced the $\gamma$-rays from the reaction $^7$Li(p,$\gamma$). Subsequent to the irradiations the decay curves of the emitted $\beta$-rays were measured. ``Als Resultat von 600 durchgef\"uhrten Bestrahlungen erhielten wir die in Fig. 13 aufgezeichnete Zerfallskurve mit einer Halbwertszeit von T = 1.06 $\pm$ 0.03 sec.'' (As a result of 600 irradiations we achieved the decay curve shown in Figure 13 with a halflife of T = 1.06 $\pm$ 0.03 sec.). This half-life agrees with the presently accepted value of 859.6(14)~ms. A previously reported halflife of 4.5~m \cite{1937Poo01} could not be confirmed .

\subsection{$^{40}$Ca}\vspace{0.0cm}

$^{40}$Ca was first observed by Dempster in 1922. He reported his result in ``Positive-ray Analysis of Potassium, Calcium and Zinc'' \cite{1922Dem01}. Positive-ray analysis was used to identify $^{40}$Ca. ``With the calcium thus prepared it was found that the component at 44 was still present, and was approximately 1/70 as strong as the main component. We therefore conclude that calcium consists of two isotopes with atomic weights 40 and 44.'' A year earlier Thomson observed a broad peak around mass 40, however, the resolution was not sufficient to determine which and how many of the isotopes 39, 40 and 41 exist \cite{1921Tho02}.

\subsection{$^{41}$Ca}\vspace{0.0cm}

The first identification of $^{41}$Ca was described in ``A Study of the Protons from Calcium under Deuteron Bombardment'' by Davidson in 1939 \cite{1939Dav01}. CaO targets were bombarded by 3.1 MeV deuterons at the Yale University cyclotron and proton absorption spectra were recorded. A group of protons with a range of 66 cm was attributed to the formation of $^{41}$Ca. ``Since calcium is predominantly Ca$^{40}$ (96.76 percent), one can almost certainly attribute this group to the reaction Ca$^{40}$+H$^2\to$Ca$^{41}$+H$^1$, giving positive evidence for the actual formation of Ca$^{41}$.''

\subsection{$^{42,43}$Ca}\vspace{0.0cm}

Aston observed $^{42,43}$Ca for the first time in 1934: ``Calcium Isotopes and the Problem of Potassium'' \cite{1934Ast02}. The relative abundances were measured with a mass spectrograph. ``Photometry gives the following provisional constitution for calcium: Mass number (Abundance): 40 (97), 42 (0.8), 43 (0.2) 44 (2.3).''

\subsection{$^{44}$Ca}\vspace{0.0cm}

$^{44}$Ca was first observed by Dempster in 1922. He reported his result in ``Positive-ray Analysis of Potassium, Calcium and Zinc'' \cite{1922Dem01}. Positive-ray analysis was used to identify $^{44}$Ca. ``With the calcium thus prepared it was found that the component at 44 was still present, and was approximately 1/70 as strong as the main component. We therefore conclude that calcium consists of two isotopes with atomic weights 40 and 44.''

\subsection{$^{45}$Ca}\vspace{0.0cm}

In the 1940 paper ``The Radioactive Isotopes of Calcium and Their Suitability as Indicators in Biological Investigations'' Walke et al. described the discovery of $^{45}$Ca \cite{1940Wal02}. Calcium was bombarded with 8 MeV deuterons at Berkeley and activated samples were placed inside a large expansion chamber. The number of positron tracks on photographs were counted over a 6 month period. A half-life of 180(10)~d was observed. ``...it is, therefore, probable that this long-lived $\beta$-radioactive calcium isotope is Ca$^{45}$ produced by the reaction: Ca$^{44}$+H$^2\to$Ca$^{45}$+H$^1$; Ca$^{45}\to$Sc$^{45}$+\textit{e}$^-$.'' The reported half life is in agreement with the currently accepted value of 162.61(9)~d.

\subsection{$^{46}$Ca}\vspace{0.0cm}

Nier reported the discovery of $^{46}$Ca in 1938 in his paper ``The Isotopic Constitution of Calcium, Titanium, Sulfur and Argon'' \cite{1938Nie01}. Calcium metal was baked in a small furnace in front of a mass spectrometer and positive ion peaks were observed at 550~$^\circ$C was used to identify $^{46}$Ca. ``One sees here, in addition to the previously known isotopes 40, 42, 43, and 44, two new peaks, one at mass 48 and one at mass 46.''

\subsection{$^{47}$Ca}\vspace{0.0cm}

In 1951 Batzel et al. described the first observation of $^{47}$Ca in ``The High Energy Spallation Products of Copper'' \cite{1951Bat01}. $^{47}$Ca was formed by spallation of copper by 340 MeV protons at the Berkeley 184-inch cyclotron. The existence of $^{47}$Ca was determined from the observation of the decay of $^{47}$Sc. ``One was the 150-day Ca$^{45}$ and the other was a 4.8$\pm$0.2-day beta-emitter with an energy of about 1.2 Mev as determined by an aluminum absorption measurement. This activity is probably the 5.8-day calcium activity reported as Ca$^{47}$ by Matthews and Pool. The growth of a 3.4-day scandium was observed in the decay of the calcium fraction and the scandium daughter was milked from the fraction.'' The 4.8(2)~d half-life is consistent with the currently accepted value of 4.536(3)~d. The mentioned activity by Matthews and Pool was only reported in a conference abstract \cite{1947Mat01}.

\subsection{$^{48}$Ca}\vspace{0.0cm}

Nier reported the discovery of $^{48}$Ca in 1938 in his paper ``The Isotopic Constitution of Calcium, Titanium, Sulfur and Argon'' \cite{1938Nie01}. Calcium metal was baked in a small furnace in front of a mass spectrometer and positive ion peaks were observed at 550~$^\circ$C was used to identify $^{48}$Ca. ``One sees here, in addition to the previously known isotopes 40, 42, 43, and 44, two new peaks, one at mass 48 and one at mass 46.''

\subsection{$^{49}$Ca}\vspace{0.0cm}

$^{49}$Ca was first observed by der Mateosian and Goldhaber in 1950, reported in ``The Question of Isomerism in Ca$^{49}$'' \cite{1950Mat01}. Enriched calcium was exposed to slow neutrons from the Argonne heavy water reactor and $\beta$-decay curves were recorded following chemical separation. ``To our surprise, we were unable to confirm the existence of either of the reported activities when Ca enriched in the isotope of mass 48 (62 percent Ca$^{48}$) was exposed to slow neutrons from the Argonne heavy water reactor. Instead, we noticed two activities of 8.5 min. and 1 hr. half-life... By chemical separation we could show that the 8.5-min. activity was due to a Ca isotope, Ca$^{49}$, and the 1-hr. activity due to a Sc isotope, Sc$^{49}$.'' This measured half-life of 8.5~m is consistent with the currently accepted value of 8.718(6)~m. The unconfirmed activities mentioned in the quote refer to half-lives of 30~m and 2.5~h reported in 1940 \cite{1940Wal02}.

\subsection{$^{50}$Ca}\vspace{0.0cm}

Shida et al. reported the discovery of $^{50}$Ca in ``New Nuclide Ca$^{50}$ and its Decay Scheme'' in 1964 \cite{1964Shi01}. Enriched calcium was bombarded by a 3.2 MeV triton beam from an electrostatic accelerator in Kawasaki. Gamma-ray spectra were measured at various times following the irradiation. ``The weighted average of the half-life is 9$\pm$2 sec. Since it was not possible to assign this activity to any known isotopes, it was suspected to be due to Ca$^{50}$... The results described above seem to be a good basis to attribute the two gamma rays to Ca$^{50}$.''  The measured half-life is in reasonable agreement with the currently accepted value of 13.9(6)~s. A previous attempt to identify $^{50}$Ca did not succeed \cite{1960Ehm01}.

\subsection{$^{51}$Ca}\vspace{0.0cm}

In 1980 Huck et al. described the first observation of $^{51}$Ca in the paper ``$\beta$ Decay of $^{51}$Ca'' \cite{1980Huc01}. $^{51}$K was produced by bombarding uranium with 600 MeV protons at the CERN synchrotron, which decayed to $^{51}$Ca through positron emission. Decay curves of $\gamma$-ray spectra were measured. ``From the decay of the six strongest lines in the multispectrum, the half-life of $^{51}$Ca was found equal to 10.0$\pm$0.8 s.'' This half-life corresponds to the currently accepted value. Only a month later Mayer et al. independently reported the detection of $^{51}$Ca by measuring the mass excess \cite{1980May01}.

\subsection{$^{52}$Ca}\vspace{0.0cm}

$^{52}$Ca was discovered by Huck et al. in 1985 and reported in ``Beta Decay of the New Isotopes $^{52}$K, $^{52}$Ca, and $^{52}$Sc; a Test of the Shell Model far from Stability'' \cite{1985Huc01}. A uranium target was fragmented by 600 MeV protons at the CERN synchrotron. Beta-decay curves and $\beta$- and $\gamma$-ray spectra were measured following online mass separation. ``A 4.6$\pm$0.3 s half-life is observed in the decay of other lines (e.g., 675, 961, 1636, and 2070 keV) and is attributed to the activity of the $^{52}$Ca parent. This assignment was confirmed by the results of separate multispectrum measurements where the decay of $^{51}$K (T$_{1/2}$=110 ms) and the growth of $^{52}$Ca (T$_{1/2}$=4.6 s) were simultaneously observed.'' The measured half-life corresponds to the currently accepted value.

\subsection{$^{53}$Ca}\vspace{0.0cm}

Langevin et al. reported the discovery of $^{53}$Ca in 1983 in ``$^{53}$K, $^{54}$K And $^{53}$Ca: Three New Neutron Rich Isotopes'' \cite{1983Lan01}. Iridium was fragmented by 10 GeV protons from the CERN synchrotron to produce neutron rich potassium isotopes, which then decayed into calcium isotopes. Neutrons were measured in coincidence with $\beta$-rays after the potassium was mass separated. ``This work gives evidence for three new K and Ca isotopes and provides further information on half-lives and P$_n$ values.'' The measured half-life of 90(15)~ms is somewhat smaller than the recent measurement of 230(60)~ms \cite{2008Man01}. Mantica et al. analyzed their data with both half-life values and found the fit for the longer half-life slightly better based on the R$^2$ regression analysis. In addition, the authors mentioned that a second $\beta$-decaying state could have contributed to the longer half-life value \cite{2008Man01}.

\subsection{$^{54-56}$Ca}\vspace{0.0cm}

$^{54}$Ca, $^{55}$Ca, and $^{56}$Ca were first observed by Bernas et al. in 1997, reported in ``Discovery and Cross-section Measurement of 58 New Fission Products in Projectile-fission of 750$\cdot$A MeV $^{238}$U'' \cite{1997Ber01}. Uranium ions were accelerated to 750 A$\cdot$MeV by the GSI UNILAC/SIS accelerator facility and bombarded a beryllium target. The isotopes produced in the projectile-fission reaction were separated using the fragment separator FRS and the nuclear charge Z for each was determined by the energy loss measurement in an ionization chamber. ``The mass identification was carried out by measuring the time of flight (TOF) and the magnetic rigidity B$\rho$ with an accuracy of 10$^{-4}$.''  11, 6 and 3 counts of $^{54}$Ca, $^{55}$Ca and $^{56}$Ca were observed, respectively.

\subsection{$^{57,58}$Ca}\vspace{0.0cm}

$^{57}$Ca and $^{58}$Ca were discovered by Tarasov et al. in 2009 and published in ``Evidence for a Change in the Nuclear Mass Surface with the Discovery of the Most Neutron-Rich Nuclei with 17 $\le$ Z $\le$ 25'' \cite{2009Tar01}. Beryllium and tungsten targets were irradiated by a 132 MeV/u $^{76}$Ge ions accelerated by the Coupled Cyclotron Facility at the National Superconducting Cyclotron Laboratory at Michigan State University. $^{57}$Ca and $^{58}$Ca were produced in projectile fragmentation reactions and identified with a two-stage separator consisting of the A1900 fragment separator and the S800 analysis beam line. ``The observed fragments include fifteen new isotopes that are the most neutron-rich nuclides of the elements chlorine to manganese ($^{50}$Cl, $^{53}$Ar, $^{55,56}$K, $^{57,58}$Ca, $^{59,60,61}$Sc, $^{62,63}$Ti, $^{65,66}$V, $^{68}$Cr, $^{70}$Mn).''

\section{Discovery of $^{98-135}$In}

Thirty-eight indium isotopes from A = $98-135$ have been discovered so far; these include 2 stable, 16 proton-rich and 20 neutron-rich isotopes.  According to the HFB-14 model \cite{2007Gor01}, $^{165}$In should be the last particle stable neutron-rich nucleus ($^{160}$In is calculated to be unbound). Along the proton dripline one more isotope is predicted to be stable and it is estimated that five additional nuclei beyond the proton dripline could live long enough to be observed \cite{2004Tho01}. Thus, there remain 35 isotopes to be discovered. About 50\% of all possible indium isotopes have been produced and identified so far and a summary is presented in Table 1.

\begin{figure}
	\centering
	\includegraphics[scale=.5]{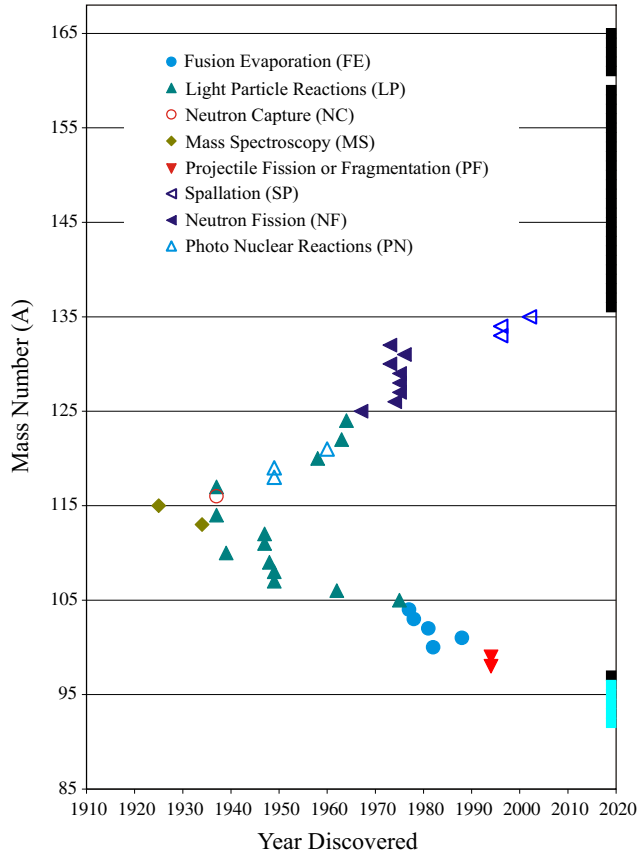}
	\caption{Indium isotopes as a function of time they were discovered. The different production methods are indicated. The solid black squares on the right hand side of the plot are isotopes predicted to be bound by the HFB-14 model.  On the proton-rich side the light blue squares correspond to unbound isotopes predicted to have lifetimes larger than $\sim 10^{-9}$~s.}
\label{f:year-in}
\end{figure}

Figure \ref{f:year-in} summarizes the year of first discovery for all indium isotopes identified by the method of discovery. The range of isotopes predicted to exist is indicated on the right side of the figure.  The radioactive indium isotopes were produced using fusion evaporation (FE), projectile fragmentation or projectile fission (PF), light-particle reactions (LP), neutron capture (NC), neutron-induced fission (NF), photo-nuclear reaction (PN) and spallation reactions (SP). The stable isotopes were identified using mass spectroscopy (MS). Heavy ions are all nuclei with an atomic mass larger than A=4 \cite{1977Gru01}. Light particles also include neutrons produced by accelerators. In the following, the discovery of each indium isotope is discussed in detail.

\subsection{$^{98,99}$In}\vspace{0.0cm}

The discovery of $^{98}$In and $^{99}$In was presented in ``Production and Identification of $^{100}$Sn'' by Schneider et al. in 1994 \cite{1994Sch01}. $^{98}$In and $^{99}$In were produced from a beryllium target bombarded by a 1095 A$\cdot$MeV $^{124}$Xe beam from the heavy-ion synchrotron SIS at GSI, Darmstadt. The products were separated with the fragment separator FRS and identified in flight by recording magnetic rigidity, multiple time-of-flights, and energy. ``The individual isotopes are clearly resolved... The majority of the events are assigned to $^{101}$Sn, the new isotope $^{99}$In, and $^{100}$In. The four events at {\it M/Q} $\sim$ 2.0 and $\Delta E \sim$ 960 a.u. in Fig. 2 are preliminarily attributed to $^{98}$In.'' 142 events of $^{99}$In were recorded.

\subsection{$^{100}$In}\vspace{0.0cm}

In 1982 the article, ``Investigations of Very Neutron-Deficient Isotopes Below $^{100}$Sn in $^{40}$Ca-Induced Reactions,'' by Kurcewicz et al. reported the discovery of $^{100}$In \cite{1982Kur01}. A 4.0 MeV/u $^{40}$Ca from the heavy-ion accelerator UNILAC at GSI was used to produce $^{100}$In in the fusion evaporation reaction $^{63}$Cu($^{40}$Ca,3n). Beta-delayed protons were measured following online mass separation. These particles were mass separated and analyzed by $\beta$- x- and $\gamma$- rays. ``From systematic considerations... the $\beta$-delayed protons observed at 97, 99 and 100 mass numbers were assigned to $^{97}$Cd, $^{99}$Cd and $^{100}$In, respectively.'' No half-life was extracted due to the limited statistics.

\subsection{$^{101}$In}\vspace{0.0cm}

In ``Decay Study of Neutron-Deficient $^{101}$In'' $^{101}$In was reported for the first time in 1988 by Huyse et al. \cite{1988Huy01}. At the Instituut voor Kern- en Stralingsfysica in Leuven a 240 MeV $^{20}$Ne beam bombarded a $^{92}$Mo target and $^{101}$In was separated and identified with the Leuven Isotope Separator On Line LISOL. ``The very neutron-deficient nucleus $^{101}$In has been identified for the first time by studying the $\beta$-delayed $\gamma$ rays of on-line mass-separated samples. The deduced half-life is 16(3)~s.'' This half-life is included in the weighted average of the current value of 15.1(3)~s.

\subsection{$^{102}$In}\vspace{0.0cm}

In 1981 Beraud et al. discovered $^{102}$In as reported in ``Identification and Decay of $^{102}$In, New Neutron Deficient Isotope Close to $^{100}$In,'' \cite{1981Ber01}. An 86 MeV $^{14}$N beam from the Grenoble cyclotron produced $^{102}$In in the fusion-evaporation reaction $^{92}$Mo($^{14}$N,4n). Gamma- and X-rays were measured following mass separation. ``Although no X-X ray characteristic of Cd element could be seen due to the presence of an enormous amount of K-X lines associated to $^{102}$Ag $\rightarrow$ Pd decay (Ag/In production ratio $>$ 10$^3$, the four lines never seen before belong necessarily to the $^{102}$In $\rightarrow$ $^{102}$Cd decay.'' The measured half-life of 24(4)~s agrees with the presently adopted value of 22(1)~s.

\subsection{$^{103}$In}\vspace{0.0cm}

The discovery of $^{103}$In was described by Lhersonneau et al. in ``Decay of neutron-deficient $^{103}$In and $^{103}$Cd Isotopes'' in 1978 \cite{1978Lhe01}. $^{14}$N was accelerated by the Louvain-la-Neuve CYCLONE cyclotron to 72 MeV and bombarded a natural molybdenum filament. $^{103}$In was produced with the fusion-evaporation reaction $^{92}$Mo($^{14}$N,3n) and separated with the online separator LISOL. The isotopes were identified by $\gamma$-ray, X-ray, and conversion electron measurements. ``The newly discovered activity $^{103}$In (T$_{1/2}$=1.08$\pm$0.11~min) was found to be populated mainly the 7/2$^{+}$ excited $^{103}$Cd state at 188~keV.'' This half-life is currently the only value measured.

\subsection{$^{104}$In}\vspace{0.0cm}

In 1977 the article ``The decay of $^{104}$In'' by Varley et al. presented the discovery of $^{104}$In \cite{1977Var01}. A $\sim$100~MeV $^{16}$O beam from the Manchester heavy-ion linear accelerator bombarded a $^{92}$Mo target to form $^{104}$In in the fusion-evaporation reaction $^{92}$Mo($^{16}$O,p3n). $^{104}$In was identified with the He-jet recoil transport system HeJRT. ``Measurements of half-lives, excitation functions gamma-x-ray and gamma-gamma coincidences have allowed the identification of gamma rays emitted in the decay of an isomer of $^{104}$In.'' The half-life of 1.5(2)~m is consistent with the current value of 1.80(3)~m for this isomer. Previously assigned half-life values of 25(6)~m and 4.6(2)~m to $^{104}$In \cite{1971Ina01} could not be confirmed.

\subsection{$^{105}$In}\vspace{0.0cm}

Rivier and Moret describe the $^{105}$In observation of 1975 in ``Mise en Evidence de L'isotope $^{105}$In et Etude de la Desintegration $^{105}$In $\rightarrow ^{105}$Cd'' \cite{1975Riv01}. Enriched $^{106}$Cd targets were bombarded with 19-31~MeV protons from the Grenoble variable energy cyclotron. $^{105}$In was produced in the (p,2n) reaction and identified by measuring $\gamma$-$\gamma$ coincidences. ``A new isotope $^{105}$In was produced by means of the reaction $^{106}$Cd(p,2n).'' The measured half-life for the ground state of 5.1(3)~m agrees with the currently adopted value of 5.07(7)~m. The article was published two years after submission. The $^{105}$In results were included in a separate article submitted two weeks after the paper by Rivier and Moret and published within six months \cite{1973Rou01}. It should also be mentioned that in 1974 another article reported the discovery of $^{105}$In \cite{1974Bur01}.

\subsection{$^{106}$In}\vspace{0.0cm}

In ``New Isotope Indium-106'' the discovery of $^{106}$In was reported in 1962 by Catura and Richardson \cite{1962Cat01}. Enriched $^{106}$Cd targets were bombarded by 14~MeV protons from the UCLA cyclotron. $^{106}$In, produced in the (p,n) charge-exchange reaction was identified by $\gamma$-ray measurements following chemical separation. ``Measurements on the yield of gamma rays above 1.8 Mev as a function of proton energy indicated the 5.3-min activity to be the result of a {\it p,n} reaction and placed an upper limit on its threshold of 8 Mev. With the above information this activity can definitely be assigned to In$^{106}$.'' The measured half-life of 5.3~m is close to the currently accepted value of 6.2(1)~m.

\subsection{$^{107,108}$In}\vspace{0.0cm}

In 1949 Mallary and Pool discovered $^{107}$In and $^{108}$In in ``Radioactive In$^{107}$, In$^{108}$, In$^{109}$ and Sn$^{108}$'' \cite{1949Mal01}. 10 MeV Deuterons and 5 MeV protons from the Mendenhall Laboratory at Ohio State University bombarded enriched $^{106}$Cd and $^{108}$Cd targets to produce $^{107}$In and $^{108}$In, respectively. Decay curves measured with a spectrometer counter and a Wulf unifilar electrometer were recorded following chemical separation. ``When cadmium enriched in isotope 106 was bombarded with deuterons and with protons, there was produced in the indium fraction a new radioactive isotope which decayed with a 33$\pm$2~min half-life by emitting positrons and gamma-rays in excess of the annihilations radiation.... The mass assignment is thus made to isotope 107 instead of 106... Two genetically related isotopes in tin and indium have been assigned to mass number 108. The indium isotope, which is produced by the decay of the tin isotope and by bombarding cadmium 108 with deuterons, decays with a half-life of about 55 min. by emitting positrons of 2-Mev energy and gamma-rays.'' These half-lives agree with the presently accepted values of 32.4(3)~m and 58.0(12)~m for $^{107}$In and $^{108}$In, respectively. A 5-h half-life had previously incorrectly been assigned to $^{108}$In \cite{1948Gho01}.

\subsection{$^{109}$In}\vspace{0.0cm}

$^{109}$In was first reported in ``Excitation Curves of ($\alpha$,n); ($\alpha$,2n); ($\alpha$,3n) Reactions on Silver'' by Ghoshal in 1948 \cite{1948Gho01}. $^{109}$In was produced by bombarding silver targets with $\alpha$-particles accelerated by the Berkeley 60-in cyclotron up to 37 MeV. The isotopes were separated with a mass-spectrograph and excitation functions and decay curves were recorded. ``The 5.2 hr. period is produced by Ag$^{107}$($\alpha$,2n)In$^{109}$ reaction. The excitation curve is similar to the excitation curve of In$^{111}$, as is expected, since both are products of ($\alpha$,2n) reactions.'' The measured half-life is close to the currently adopted value of 4.167(18)~h.

\subsection{$^{110}$In}\vspace{0.0cm}

In the 1939 article, ``Proton Activation of Indium and Cadmium,'' Barnes reported the first observation of $^{110}$In \cite{1939Bar01}. Cadmium foils were bombarded by 7.2 MeV protons from the University of Rochester's cyclotron and decay curves were measured with an ionization chamber. ``The positron activity with half-life of 65$\pm$5~min. has not been previously reported... In$^{106}$, In$^{108}$ and In$^{110}$ must be positron emitters, and since Cd$^{110}$ is ten times as abundant as either Cd$^{106}$ or Cd$^{108}$ this activity is tentatively assigned to In$^{110}$.'' This half-life agrees with the value of the 69.1(5)~m isomeric state.

\subsection{$^{111,112}$In}\vspace{0.0cm}

``The Radioactive Indium Isotopes of Mass Numbers 111 and 112'' by Tendam and Bradt was published in 1947 identifying $^{111}$In and $^{112}$In \cite{1947Ten01}. At Purdue University silver targets were bombarded with 15-20 MeV $\alpha$-particles. Indium was identified by chemical analysis, and the isotopes were identified via excitation energy measurements and decay curves. ``It is seen from its excitation curve that the 2.7-day period is the product of an ($\alpha$,2n) reaction with a threshold of 15.5$\pm$0.5 MeV and must be assigned to In$^{111}$... Since its excitation curve is almost identical with that of the 66-min. In$^{110}$, produced by the Ag$^{107}$($\alpha$,n)In$^{110}$ reaction, the 23-min. period must be assigned to mass number 112 as the product of the Ag$^{109}$($\alpha$,n)In$^{112}$ reaction.'' These half-lives are consistent with the currently accepted values of 2.8047(4)~d and 20.56(6)~m, for $^{111}$In and $^{112}$In, respectively. The half-life for $^{112}$In corresponds to an isomeric state. Lawson and Cork had previously assigned a $\sim$20~m half-life to $^{111}$In in several papers \cite{1937Law01,1939Cor01,1940Law01}. Barnes also assigned an 18-20~m half-life to $^{111}$In and he attributed a 2.7~d half-life to an $^{112}$In isomer in 1939 \cite{1939Bar01}. Cork and Lawson assigned a 65.0(45)~h half-life first to $^{113}$In \cite{1939Cor01} and later to $^{112}$In \cite{1940Law01}.

\subsection{$^{113}$In}\vspace{0.0cm}

Wehrli reported the discovery of $^{113}$In in the 1934 article ``Das Indium-Isotop 113'' \cite{1934Weh01}. $^{113}$In was identified by means of anode ray spectrography. ``Gemeinsam mit E. Meischer habe ich im Bandenspektrum des InJ 2 schwache Kanten festgestellt, welche als Isotopenkanten gedeutet und dem In$_{113}$J zugeordnet wurden.'' (Together with E. Meischer I have determined two weak edges in the line spectrum of InJ, which were interpreted as isotope edges and assigned to $^{113}$In.) Further details were presented in a subsequent publication \cite{1934Weh02}.

\subsection{$^{114}$In}\vspace{0.0cm}

The isotope $^{114}$In was first identified in 1937 by Lawson and Cork in``The Radioactive Isotopes of Indium'' \cite{1937Law01}. Indium was irradiated with 14 to 20 MeV neutrons produced from the bombardment of lithium with 6.3 MeV deuterons at the University of Michigan. $^{114}$In was identified via decay curve measurements. ``The 50-day period has so far been observe only when the activation has been with fast neutrons. This therefore might be placed as either an isomer of 112 or 114. It has been tentatively placed as 114.'' The 50~d half-life agrees with the currently accepted value for the 49.51(1)~d isomeric state. The half-life of the $^{114}$In ground state is 71.9(1)~s. Half-lives of 1~m \cite{1937Cha01} and 1.1~m \cite{1937Poo01} had been measured previously but no mass assignments were made. Also, a 13~s half-life had been assigned incorrectly to $^{114}$In \cite{1937Cor01}.

\subsection{$^{115}$In}\vspace{0.0cm}

Aston described the discovery of $^{115}$In in the 1925 article ``The Mass Spectra of Chemical Elements, Part VI. Accelerated Anode Rays Continued'' \cite{1924Ast01}. $^{115}$In was detected using the accelerated anode ray method with a solution of hydrofluoric acid: ``This incorporated into the anode gave a mass spectrum showing one line only at 115.''

\subsection{$^{116}$In}\vspace{0.0cm}

The isotope $^{116}$In was first identified in 1937 by Lawson and Cork in``The Radioactive Isotopes of Indium'' \cite{1937Law01}. Indium was irradiated with slow neutrons at the University of Michigan. Decay curves of $\beta$-activity were measured and half-lives extracted, ``...although the 13-second and 54-minute periods could have been associated with either 114 or 116 the are undoubtedly due to 116.'' The 13~s half-life agrees with the currently adopted value for the ground state of 14.10(3)~s. The 54~m half-life corresponds to the 54.29(13)~m isomeric state. The 13~s and the 54~m had been previously observed but without a definite mass assignment \cite{1935Ama01}. In an article published a few months earlier Cork and Thornton had associated a 58~m half-life with $^{116}$In, however, without an actual measurement \cite{1937Cor01}.

\subsection{$^{117}$In}\vspace{0.0cm}

The discovery of $^{117}$In was described in 1937 by Cork and Thornton in the article ``The Disintegration of Cadmium with Deuterons'' \cite{1937Cor01}. A 6.3 MeV deuteron beam bombarded metallic cadmium at the University of Michigan.  Indium was successively abstracted from chemically separated cadmium and decay curves measured. ``The long-period cadmium activity gives rise to a radioactive indium of half-life 2.3 hr.'' This half-life was assigned to $^{117}$In in a table and is consistent with the currently adopted value of the 116.2(3)~m isomeric state.

\subsection{$^{118,119}$In}\vspace{0.0cm}

In 1949 $^{118}$In and $^{119}$In were first observed by Duffield and Knight in ``In$^{118}$ and In$^{119}$ produced by Photo-Disintegration of Tin'' \cite{1949Duf02}. At the University of Illinois, 23 MeV X-rays bombarded enriched $^{119}$Sn and $^{120}$Sn to produce $^{118}$In and $^{119}$In, respectively. Decay curves were recorded which in the case of $^{119}$In was preceded by chemical separation. ``An examination of the indium activities produced by the irradiation of tin with 23 MeV betatron x-rays at this laboratory has led to the identification of two additional periods which can be assigned to In$^{118}$ and In$^{119}$ on the basis of evidence outlined below.'' The measured half-lives for $^{118}$In (4.5(5)~m) and $^{119}$In (17.5(10)~m) agree with the currently accepted values for isomeric states in these nuclei of 4.364(7)~m and 18.0(3)~m, respectively.

\subsection{$^{120}$In}\vspace{0.0cm}

In ``Radioactivity of In$^{120}$ and Sb$^{120}$'' McGinnis reported the discovery of $^{120}$In in 1958 \cite{1958McG01}. $^{120}$In was produced in a (n,p) charge-exchange reaction by bombarding natural tin with 20 MeV neutrons. No chemical separation was performed and $\gamma$-rays were measured with a scintillation detector. ``The data of Table VII are the basis for assigning the ~55~s activity to In$^{120m}$.'' This half-life is consistent with either of two isomeric states with half-lives of 47.3(5)~s and 46.2(8)~s.

\subsection{$^{121}$In}\vspace{0.0cm}

In 1960 Yuta and Morinaga identified $^{121}$In for the first time in ``Study of Heavy Odd-Mass Indium Isotopes from the ($\gamma$,p) Reaction on Tin'' \cite{1960Yut01}. Targets of enriched $^{122}$SnO$_{2}$ were bombarded by 25 MeV bremsstrahlung from the 25-MeV betatron at Tohoku University. Gamma-ray spectra were measured with a 4''$\times$4'' NaI crystal and $\beta$ decay curves were recorded. ``Here, a new peak at 0.94 MeV is clearly seen. This peak decayed with a half-life of 30$\pm$3 sec... it is assigned to the g$_{9/2}$ state of In$^{121}$.'' This half-life is consistent with the currently accepted value of 23.1(6)~s. Previously reported half-lives of 12~m and 32~m \cite{1957Nus01} could not be confirmed.

\subsection{$^{122}$In}\vspace{0.0cm}

The discovery of $^{122}$In by Kantele and Karras was reported in the 1963 publication ``New Isotope In$^{122}$'' \cite{1963Kan01}. A 14-15 MeV beam of neutrons bombarded a $^{122}$Sn enriched target at the University of Arkansas 400 kV Cockcroft-Walton accelerator and produced $^{122}$In in the (n,p) charge exchange reaction. $\gamma$- and $\beta$-radiation and $\gamma$-$\gamma$ coincidences were measured. ``In connection with a systematic study of the level structure of even tin isotopes resulting from the decay of neutron-excess indium isotopes, a new 7.5-sec activity was found and was assigned to the hitherto unknown isotope In$^{122}$.'' This 7.5(8)~s half-life could be either one of two isomeric states of 10.3(6)~s or 10.8(4)~s.

\subsection{$^{123}$In}\vspace{0.0cm}

In 1960 Yuta and Morinaga identified $^{123}$In for the first time in ``Study of Heavy Odd-Mass Indium Isotopes from the ($\gamma$,p) Reaction on Tin'' \cite{1960Yut01}. Targets of enriched $^{124}$SnO$_{2}$ were bombarded by 25 MeV bremsstrahlung from the 25-MeV betatron at Tohoku University. Gamma-ray spectra were measured and $\beta$ decay curves were recorded. ``A peak at 1.10 MeV appears here and has a half-life of 10 sec. It is assigned to the g$_{9/2}$ state of $^{123}$In as in the cases of Sn$^{120}$ and Sn$^{122}$.'' The measured half-life of 10(2)~s is close to the currently accepted value of 6.17(5)~s.

\subsection{$^{124}$In}\vspace{0.0cm}

In 1964 the article ``New Isotope In$^{124}$,'' by Karras reported the discovery of $^{124}$In \cite{1964Kar01}. The neutron generator at the University of Arkansas provided 14-15 MeV neutrons which bombarded enriched $^{124}$Sn and produced $^{124}$In in the (n,p) charge exchange reaction. $\beta$- and $\gamma$-ray spectra were measured. ``Irradiation of Sn$^{124}$ samples with 14-15 MeV neutrons was found to produce a new radioactive nuclide which was assigned to In$^{124}$.'' The measured 3.6~s half-life agrees with the currently adopted value of 3.12(9)~s for the ground state or with the 3.7(2)~s isomeric state.

\subsection{$^{125}$In}\vspace{0.0cm}

In ``Short-Lived Fission Products'' the first observation of $^{125}$In was reported in 1967 by Fritze and Griffiths \cite{1967Fri01}. $^{125}$In was produced via neutron induced fission of $^{235}$U at the McMaster University Reactor. The isotope was identified by its daughter activity following chemical separation. ``Proof of the presence of a given nuclide depended on the identification of a known daughter activity resulting from the decay of the unknown short-lived parent, which had been separated completely from daughter activities as soon as possible after the end of the irradiation... Starting 13 min after the end of the irradiation $\gamma$-spectra were taken at 7 min intervals and showed the presence of 40 min $^{123}$Sn (160 keV) and 10 min $^{125m}$Sn (335 keV).''

\subsection{$^{126}$In}\vspace{0.0cm}

Grapengiesser et al. reported the observation of $^{126}$In in ``Survey of Short-lived Fission Products Obtained Using the Isotope-Separator-On-Line Facility at Studsvik'' in 1974 \cite{1974Gra01}. $^{126}$In was produced by neutron induced fission and identified at the OSIRIS isotope-separator online facility at the Studsvik Neutron Research Laboratory in Nyk\"oping, Sweden. In the long table of experimental half-lives of many different isotopes the half-life of $^{126}$In is quoted as 1.53(1)~s. This value is included in the currently adopted average 1.53(1)~s.

\subsection{$^{127-129}$In}\vspace{0.0cm}

In 1975 the first identification of $^{127}$In, $^{128}$In, and $^{129}$In was reported by Aleklett et al. ``Beta-Decay Properties of Strongly Neutron-Rich Nuclei'' \cite{1975Ale01}. The isotopes were produced by thermal-neutron induced fission of $^{235}$U and identified using Studsvik's OSIRIS separator. In Table 2 listing heavy fission fragments from silver to lanthanum, $^{127}$In, $^{128}$In, and $^{129}$In are identified with half-lives as quoted from the report by Grapengiesser \cite{1974Gra01}. Grapengiesser did not uniquely assign the elements; for $^{127}$In and $^{128}$In cadmium or indium are listed as possible element and no element assignment was made for $^{129}$In. While the half-lives for $^{128}$In (0.80(3)~s) and $^{129}$In (0.8(3)~s) agree, Grapengiesser quotes two values for mass 127 (1.3(2)~s and 3.7(1)~s) while Aleklett et al. quotes only a value of 3.1~s. The half-life of 3.1~s for $^{127}$In agrees with the current measurement of 3.67(4)~s for an isomeric state, the half-lives of 0.8~s agree with the currently adopted values of 0.84(6)~s and 0.61(1)~s for $^{128}$In and $^{129}$In, respectively.

\subsection{$^{130}$In}\vspace{0.0cm}

``Excited States in the Two-Neutron-Hole Nucleus $^{130}_{50}$Sn$_{80}$ Observed in the 0.53 sec $\beta^-$ Decay of $^{130}$In'' described the first observation of $^{130}$In by Kerek et al. in 1973 \cite{1973Ker01}. $^{130}$In was produced by neutron induced fission of $^{235}$U at Studsvik, Sweden, and identified utilizing the OSIRIS separator. ``Among the high-energy $\beta$-rays a short-lived component with the half-life 0.53$\pm$0.05 sec could be observed. Since the E$_{\beta^-}$ threshold exceeds all other E$_{\beta^-}$ in the chain, the half-life is assigned to the $^{130}$In$\rightarrow^{130}$Sn decay.'' This half-life value is included in the calculation of the currently adopted value of 0.54(1)~s.

\subsection{$^{131}$In}\vspace{0.0cm}

$^{131}$In was discovered by Lund and Rundstam in 1976 as reported in ``Delayed-neutron activities produced in fission: Mass range 122-146'' \cite{1976Lun01}. $^{131}$In was produced via neutron fission in a uranium target at the Studsvik R2-0 reactor and separated with the OSIRIS on-line mass-separator facility. 30 $^3$He neutron counters were used to measure the delayed neutron activities. ``The 0.29 sec activity is to be attributed to $^{131}$In for which the $\beta$ half-life has been determined to be 0.27$\pm$0.02 sec.'' This 0.29(1)~s activity agrees with the currently accepted value of 0.28(3)~s. The cited value of 0.27(2)~s referred to a ``to be published article'' by De Geer et al..

\subsection{$^{132}$In}\vspace{0.0cm}

The discovery of $^{132}$In was described in the 1973 article ``The First Excited State in the Doubly-Closed-Shell Nucleus $^{132}$Sn Populated in the 0.12 s $\beta$$^{-}$-Decay of $^{132}$In'' by Kerek et al. \cite{1973Ker02}. $^{132}$In was produced by neutron induced fission of $^{235}$U at Studsvik, Sweden, and identified utilizing the OSIRIS separator. ``A 0.12$\pm$0.02~s beta activity assigned to the decay of $^{132}$In and populating an excited state of 4041$\pm$2 keV in the doubly-closed-shell nucleus $^{132}_{50}$Sn$_{82}$ has been observed.'' This half-life is near the currently accepted value of 0.207(6)~s.

\subsection{$^{133,134}$In}\vspace{0.0cm}

In 1996 Hoff et al. reported the discovery of $^{133}$In and $^{134}$In in ``Single-Neutron States in $^{133}$Sn'' \cite{1996Hof01}. 1~GeV protons induced fission of uranium carbide at the CERN PS-Booster. Mass separation and $\beta$- and $\gamma$-decay spectroscopy was performed at the ISOLDE facility. Decay characteristics of of $^{133}$In were measured but the half-life was not extracted and assumed to be known: ``Some of the present authors attempted to determine the structure of $^{133}$Sn at the ISOLDE facility at the CERN SC, more than a decade ago... Although two $\beta$-decay states g$^{-1}_{9/2}$ and p$^{-1}_{1/2}$, were expected, only one half-life of 180$\pm$15~ms was observed.'' No specific reference is given but most likely it referred to a 1981 conference proceeding by Blomqvist et al. \cite{1981Blo01}. $^{133}$Sn was also populated by $\beta$-delayed neutron emission from $^{134}$In. ``Some distinct transitions in $^{133}$Sn clearly visible, in particular, those at 854, 1561, and 2005 keV. An analysis of their time dependence with respect to the beam pulses gave the half-life of $^{134}$In as 138$\pm$8~ms.'' The quoted half-life for $^{133}$In agrees with the currently accepted half-life of 165(3)~ms and the measured half-life for $^{134}$In is included in the currently adopted weighted average of 140(4)~ms.

\subsection{$^{135}$In}\vspace{0.0cm}

The 2002 article, ``Selective Laser Ionization of N $\ge$ 82 Indium Isotopes: The New r-process Nuclide 135In,'' by Dillmann et al. discussed the first identification of $^{135}$In \cite{2002Dil01}. A tantalum converter was bombarded with 1.4 GeV protons at the CERN ISOLDE facility. Neutrons from the converter induced fission in an adjacent UC$_x$/graphite target.$^{135}$In was separated and identified using laser ionization. ``With 92(10)~ms $^{135}$In, a new r-process nuclide has been identified...'' This half-life is currently the only measured value.

\section{Discovery of $^{100-137}$Sn}

Thirty-eight tin isotopes from A = $100-137$ have been discovered so far; these include 10 stable, 13 proton-rich and 15 neutron-rich isotopes.  According to the HFB-14 model \cite{2007Gor01}, $^{176}$Sn should be the last particle stable neutron-rich nucleus (the odd mass isotopes $^{175}$Sn and $^{173}$Sn are calculated to be unbound). Along the proton dripline two more isotopes are predicted to be stable and it is estimated that six additional nuclei beyond the proton dripline could live long enough to be observed \cite{2004Tho01}. Thus, there remain about 47 isotopes to be discovered. About 45\% of all possible tin isotopes have been produced and identified so far.

\begin{figure}
	\centering
	\includegraphics[scale=.5]{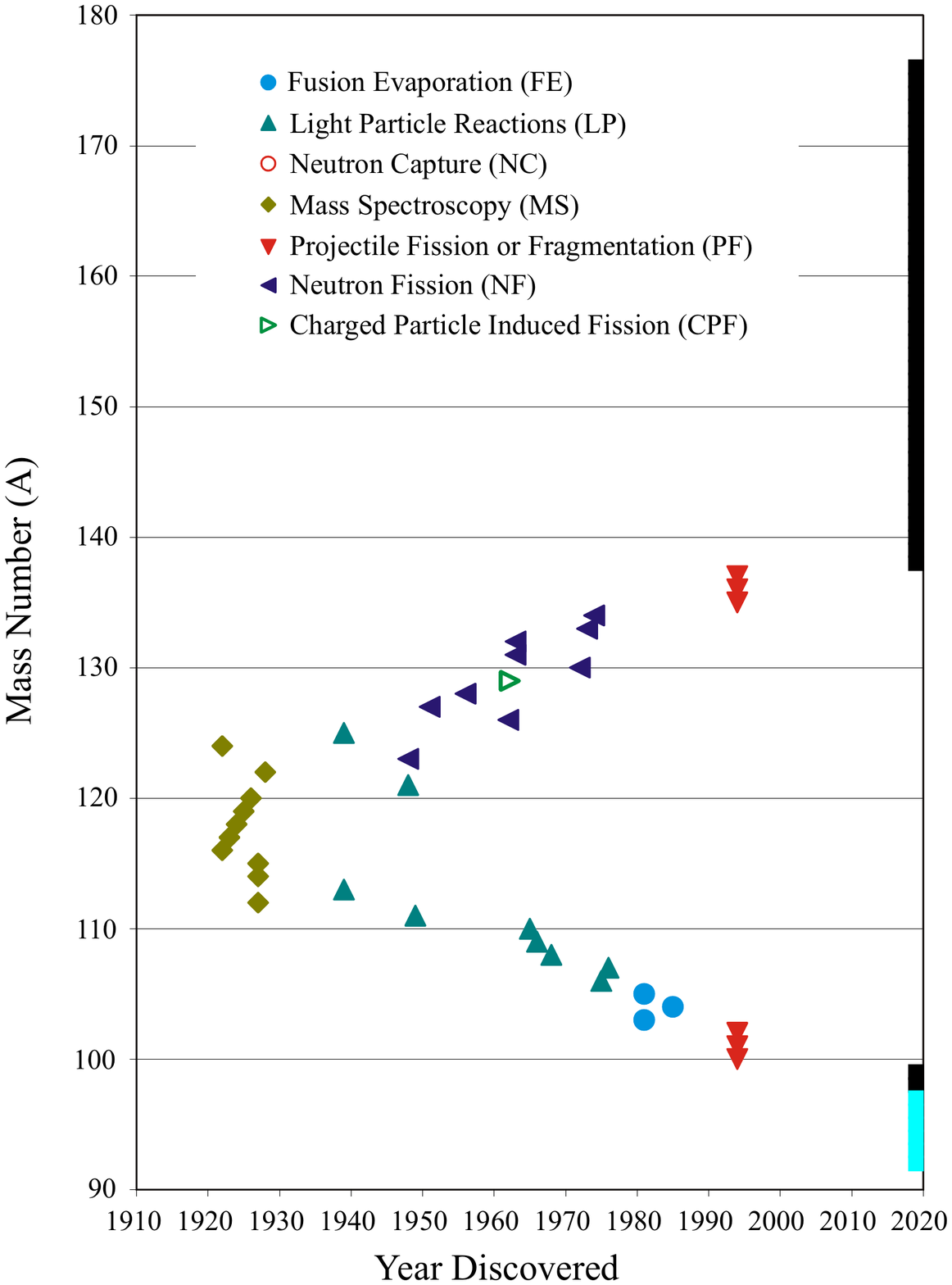}
	\caption{Tin isotopes as a function of time they were discovered. The different production methods are indicated. The solid black squares on the right hand side of the plot are isotopes predicted to be bound by the HFB-14 model.  On the proton-rich side the light blue squares correspond to unbound isotopes predicted to have lifetimes larger than $\sim 10^{-9}$~s.}
\label{f:year-sn}
\end{figure}

Figure \ref{f:year-sn} summarizes the year of first discovery for all tin isotopes identified by the method of discovery.  The range of isotopes predicted to exist is indicated on the right side of the figure.  The radioactive tin isotopes were produced using heavy-ion fusion evaporation (FE), projectile fragmentation or projectile fission (PF), light-particle reactions (LP), neutron-induced fission (NF), and charged-particle induced fission (CPF). The stable isotopes were identified using mass spectroscopy (MS). Heavy ions are all nuclei with an atomic mass larger than A=4 \cite{1977Gru01}. Light particles also include neutrons produced by accelerators. In the following, the discovery of each tin isotope is discussed in detail and a summary is presented in Table 1.

\subsection{$^{100,101}$Sn}\vspace{0.0cm}

The discovery of $^{100}$Sn and $^{101}$Sn was presented in ``Production and Identification of $^{100}$Sn'' by Schneider et al. in 1994 \cite{1994Sch01}. $^{100}$Sn and $^{101}$Sn were produced from a beryllium target bombarded by a 1095 A$\cdot$MeV $^{124}$Xe beam from the heavy-ion synchrotron SIS at GSI, Darmstadt. The products were separated with the fragment separator FRS and identified in flight by recording magnetic rigidity, multiple time-of-flights, and energy. ``The individual isotopes are clearly resolved. We attribute 7 events to the isotope $^{100}$Sn. The majority of the events are assigned to $^{101}$Sn, the new isotope $^{99}$In, and $^{100}$In.'' 70 events of $^{101}$Sn were recorded. It should be mentioned that Lewitowicz \text{et al.} \cite{1994Lew01} submitted their observation of $^{100}$Sn less than two months after Schneider et al..

\subsection{$^{102}$Sn}\vspace{0.0cm}

The discovery of $^{102}$Sn was reported in ``Identification of the Doubly-Magic Nucleus $^{100}$Sn in the reaction $^{112}$Sn+$^{nat}$Ni at 63MeV/nucleon'' by Lewitowicz et al. in 1994 \cite{1994Lew01}. A beam of 63 MeV/nucleon $^{112}$Sn bombarded a nickel target at GANIL and $^{102}$Sn was separated and identified using the Alpha and LISE3 spectrometers. ``It is then possible to calculate for a group of events, selected on the basis of the $Z$ and $A/Q$, the masses of the individual ions from the measured TKE and TOF. The resulting mass distributions for $^{104}$Sn$^{+50}$, $^{102}$Sn$^{+49}$, $^{100}$Sn$^{+48}$ and $^{105}$Sn$^{+50}$, $^{103}$Sn$^{+49}$, $^{101}$Sn$^{+48}$, are given in Figs. 2c and 2d respectively.'' Schneider et al. showed events of $^{102}$Sn in the paper submitted two months earlier, however, they only mentioned that it was strongly suppressed at the FRS setting for $^{100}$Sn \cite{1994Sch01}.

\subsection{$^{103}$Sn}\vspace{0.0cm}

In 1981 the first observation of $^{103}$Sn was described in ``The New Beta-Delayed Proton Precursors $^{103}$Sn and $^{105}$Sn'' by Tidemand-Petersson et al. \cite{1981Tid01}. The UNILAC at GSI Darmstadt was used to produce $^{103}$Sn in fusion-evaporation reactions with a 290 MeV $^{58}$Ni beam and separated with a FEBIAD ion source with a graphite catcher. ``Using $^{58}$Ni+$^{50}$Cr and $^{58}$Ni+$^{54}$Fe reactions and on-line mass separation, the new isotopes $^{103}$Sn and $^{105}$Sn with half-lives of 7$\pm$3~s and 31$\pm$6~s, respectively, were identified via their beta-delayed proton decays.'' The measured half-life for $^{103}$Sn agrees with the currently accepted value of 7.0(2)~s.

\subsection{$^{104}$Sn}\vspace{0.0cm}

In the 1985 article ``First Identification of $\gamma$ Rays in the $\beta^+$/EC decay of $^{104,105}$Sn'' Deneffe et al. reported the observation of $^{104}$Sn \cite{1985Den01}. $^{104}$Sn was formed with the CYCLONE cyclotron at Louvain-la-Neuve using 210 MeV $^{20}$Ne beam particles of 210 MeV bombarding a $^{92}$Mo target. The isotopes were separated and identified with the LISOL isotope separator. ``The $\beta^{+}$/EC decay of mass-separated $^{104}$Sn and $^{105}$Sn isotopes was studied by x-ray and $\gamma$-ray singles, as well as by x$-\gamma$ and $\gamma-\gamma$ coincidences.'' The measured half-life of 23(2)~s agrees with the accepted half-life of 20.8(5)~s.

\subsection{$^{105}$Sn}\vspace{0.0cm}

In 1981 the first observation of $^{105}$Sn was described in ``The New Beta-Delayed Proton Precursors $^{103}$Sn and $^{105}$Sn'' by Tidemand-Petersson et al. \cite{1981Tid01}. The UNILAC at GSI Darmstadt was used to produce $^{105}$Sn in fusion-evaporation reactions with a 290 MeV $^{58}$Ni beam and separated with a FEBIAD ion source with a graphite catcher. ``Using $^{58}$Ni+$^{50}$Cr and $^{58}$Ni+$^{54}$Fe reactions and on-line mass separation, the new isotopes $^{103}$Sn and $^{105}$Sn with half-lives of 7$\pm$3~s and 31$\pm$6~s, respectively, were identified via their beta-delayed proton decays.'' The measured half-life for $^{105}$Sn agrees with the currently accepted value of 34(1)~s.

\subsection{$^{106}$Sn}\vspace{0.0cm}

Burminskii et al. reported the discovery of $^{106}$Sn in the 1975 article ``A New Tin Isotope-$^{106}$Sn'' \cite{1975Bur01}. A $^{3}$He beam accelerated to 21-59 MeV by the isochronous cyclotron at the Kazakh Academy of Sciences bombarded an enriched $^{106}$Cd target. $^{106}$Sn was formed in the charge-exchange reaction $^{106}$Cd($^3$He,3n) and identified by excitation functions and $\gamma$-ray measurements. ``In addition to the $\gamma$ lines of the known isotopes we observe in the spectra $\gamma$ rays... with a half-life T$_{1/2}$=1.9$\pm$0.3~min, which we ascribe, on the basis of the identification described below, to the decay of a new isotope, $^{106}$Sn.'' This half-life agrees with the currently adopted value of 115(5)~s.

\subsection{$^{107}$Sn}\vspace{0.0cm}

The first correct identification of $^{107}$Sn was reported by Hseuh and Macias in the 1976 publication ``Identification of 2.90-min $^{107}$Sn and 50-sec $^{107}$In$^m$'' \cite{1976Hse01}. $^{107}$Sn was produced by bombarding enriched $^{105}$CdO with 30 MeV $^3$He from the Washington University cyclotron. ``$^{107}$Sn and $^{107}$In$^m$, produced in $^3$He bombardments of $^{106}$Cd and transported with a He jet system, have been identified with Ge(Li) $\gamma$-ray detectors. Half-lives of these nuclides were determined to be 2.90$\pm$0.05 min and 50.4$\pm$0.6 sec, respectively.'' This half-life is the currently accepted value. A previously measured half-life of 1.3(3)~m \cite{1972Riv01} could not be confirmed. Also the energies of two $\gamma$-rays assigned to $^{107}$Sn \cite{1969Yam01} could not be reproduced.

\subsection{$^{108}$Sn}\vspace{0.0cm}

In 1968 Yamazaki et al. observed $^{108}$Sn in ``Level and Isomer Systematics in Even Sn Isotopes'' \cite{1968Yam01}. The Berkeley 88-in. cyclotron was used to bombard enriched metallic cadmium targets with 28-50 MeV $\alpha$ particles. The first three excited states in $^{108}$Sn were detected with a Ge(Li) detector. ``Levels of even Sn isotopes (A=108-118) have been studied in Cd($\alpha,xn)$ reactions.'' A previously measured half-life of 4.5~h \cite{1949Mal01} was later not confirmed.

\subsection{$^{109}$Sn}\vspace{0.0cm}

The first observation of $^{109}$Sn was reported in the 1965 article ``Decay Scheme For Sn$^{109}$'' by Khulelidze et al. \cite{1966Khu01}. $^{109}$Cd was produced by bombarding an enriched $^{106}$Cd target with 21 MeV $\alpha$-particles at a cyclotron in Dubna, Russia. $\gamma$-rays and $\beta$-particles were recorded with a $\beta$ spectrometer and a scintillation spectrometer. ``Two lines with E$_{e} = 625$ and 648 KeV appear in the conversion electron spectrum. These lines fell off in intensity with a period of 19$\pm$2 min.'' This half-life agrees with the currently accepted value of 18.0(2)~m. The references quoted in the paper refer to conference abstracts \cite{1955Pet01,1956Pet01}.

\subsection{$^{110}$Sn}\vspace{0.0cm}

$^{110}$Sn was first observed by Bassani et al. in 1965 as reported in ``(p,t) Ground-State L=0 Transitions in the Even Isotopes of Sn and Cd at 40 MeV, N = 62 to 74'' \cite{1965Bas01}. 40 MeV protons accelerated by the University of Minnesota linear accelerator bombarded isotopic $^{112}$Sn foil targets. $^{110}$Sn was produced in the (p,t) reaction and was identified by measuring tritons with an array of eight plastic scintillators in the focal plane of 40~in. 180$^\circ$ magnetic spectrometer. The angular distribution for the ground state transition was measured between 7$^\circ$ and 25$^\circ$. The correct half-life for $^{110}$Sn was only reported two years later by Bogdanov et al. as 4(1)~h \cite{1967Bog01}. Bogdanov did not consider this a new measurement quoting a value of 4.1~h from a 1963 compilation \cite{1963Dzh01} which was based on data published in a conference abstract \cite{1955Mea01} and a thesis \cite{1956Mea01}. In 1949 a 4.5~h half-life had been incorrectly assigned \cite{1949Mal01} - and subsequently been apparently confirmed \cite{1950Lin01,1951McG01} - to $^{108}$Sn.

\subsection{$^{111}$Sn}\vspace{0.0cm}

In 1949 $^{111}$Sn was discovered by Hinshaw and Pool in ``Radioactive Tin 111'' \cite{1949Hin01}. A beam of 20 MeV $\alpha$ particles bombarded an enriched cadmium metal target at Ohio State University. $^{111}$Sn was identified by measuring decay curves with a Geiger counter in a magnetic field following chemical separation. ``The results showed that the activity was clearly obtained from the Cd$^{108}$ isotope but not from the Cd$^{106}$ or Cd$^{110}$ isotopes. The assignment of the 35-minute activity is thus made to Sn$^{111}$.''  This 35.0(5)~m half-life is included in the currently accepted average value of 35.3(6)~m.

\subsection{$^{112}$Sn}\vspace{0.0cm}

In ``Atoms and their Packing Fractions,'' published in 1927, Aston reported the discovery of $^{112}$Sn \cite{1927Ast02}. The isotopes were identified with help of a new mass spectrograph at the Cavendish Laboratory. The tin isotopes are shown in mass spectrum X of Figure 1: ``X- (a) and (b) spectra showing the even spacing of the tin monomethide and xenon lines. (c) The same with long exposure showing eleven isotopes of tin.'' $^{112}$Sn, $^{114}$Sn and $^{115}$Sn are listed in Table 1 as the tin isotopes with the weakest intensity.

\subsection{$^{113}$Sn}\vspace{0.0cm}

Livingood and Seaborg reported in 1939 the observation of $^{113}$Sn in the article ``New Periods of Radioactive Tin'' \cite{1939Liv01}. 5 MeV deuterons bombarded tin targets at the Berkeley Radiation Laboratory and radioactive decay curves were recorded. ``The only unstable tin isotope common to the reactions (Sn,dp)Sn and Cd($\alpha$,n)Sn is Sn$^{113}$. A chemical separation of tin after activation of cadmium with 16-MeV helium ions does in fact give a precipitate which contains an activity with a half-life of about 70 days (sign unknown). This is additional evidence that Sn$^{113}$ has this period and perhaps decays by K-electron capture to stable In$^{113}$.'' This half-life is in reasonable agreement with the currently adopted value of 115.09(3)~d. Only three months later Barns reported a half-life of 105~d for $^{113}$Sn \cite{1939Bar01}.

\subsection{$^{114,115}$Sn}\vspace{0.0cm}

In ``Atoms and their Packing Fractions,'' published in 1927, Aston reported the discovery of $^{114}$Sn and $^{115}$Sn \cite{1927Ast02}. The isotopes were identified with help of a new mass spectrograph at the Cavendish Laboratory. The tin isotopes are shown in mass spectrum X of Figure 1: ``X- (a) and (b) spectra showing the even spacing of the tin monomethide and xenon lines. (c) The same with long exposure showing eleven isotopes of tin.'' $^{112}$Sn, $^{114}$Sn and $^{115}$Sn are listed in Table 1 as the tin isotopes with the weakest intensity.

\subsection{$^{116-120}$Sn}\vspace{0.0cm}

$^{116}$Sn, $^{117}$Sn, $^{118}$Sn, $^{119}$Sn, and $^{120}$Sn were discovered by Aston in ``The Isotopes of Tin'' in 1922 \cite{1922Ast02}. The tin isotopes were identified with ``Half Tone'' plates installed at the Cavendish Laboratory mass spectrograph. ``Tin tetramethide was employed, and a group of eight lines corresponding approximately to atomic weights 116(c), 117(f), 118(b), 119(e), 120(a), 121(h), 122(g), 124(d) was definitely proved to be due to tin.'' The letters following the masses indicate the ordering of the observed intensity. The observation of the weakest isotope ($^{121}$Sn) proved to be incorrect.

\subsection{$^{121}$Sn}\vspace{0.0cm}

Lindner and Perlman discovered $^{121}$Sn in 1948 in ``Neutron Deficient Isotopes of Tellurium and Antimony'' \cite{1948Lin01}. An 18 MeV deuteron beam accelerated by the Berkeley 60-inch cyclotron bombarded an isotopically enriched $^{120}$Sn target and the $\beta$-decay of the tin precipitate was recorded. ``The tin fraction was found to contain a single $\beta^-$-activity of 28-hour half-life. The maximum $\beta^-$-energy was found from absorption in beryllium and from beta-ray spectrometer measurements to be about 0.4 Mev and no $\gamma$-radiation was present. The 26-hr. Sn of Livingood and Seaborg is therefore Sn$^{121}$, formed in this case by the reaction Sn$^{120}$(d,p)Sn$^{121}$.'' The 28~h half-life agrees with the currently adopted value of 27.3(4)~h. Originally $^{121}$Sn had incorrectly been observed to be stable \cite{1922Ast02}. A 24~h activity had previously been assigned to either $^{121}$Sn, $^{122}$Sb, or $^{124}$Sb \cite{1936Liv01}, and a 28(2)~h half-life to $^{113}$Sn, $^{121}$Sn, or $^{123}$Sn \cite{1936Liv02}. The 26-h activity referred to by Lindner and Perlman was not believed to belong to $^{121}$Sn by Livingood and Seaborg: ``the 26-hour tin isotope from tin plus deuterons plus neutrons is not necessarily associated with Sn$^{121}$, although it may be.'' \cite{1939Liv01}. Finally, in 1947 Seren assigned a 26~h half-life to $_{50}$Sn$^{<125}$ \cite{1947Ser01}.

\subsection{$^{122}$Sn}\vspace{0.0cm}

$^{122}$Sn was discovered by Aston in ``The Isotopes of Tin'' in 1922 \cite{1922Ast01}. The tin isotopes were identified with ``Half Tone'' plates installed at the Cavendish Laboratory mass spectrograph. ``Tin tetramethide was employed, and a group of eight lines corresponding approximately to atomic weights 116(c), 117(f), 118(b), 119(e), 120(a), 121(h), 122(g), 124(d) was definitely proved to be due to tin.'' The letters following the masses indicate the ordering of the observed intensity. The observation of the weakest isotope ($^{121}$Sn) proved to be incorrect.

\subsection{$^{123}$Sn}\vspace{0.0cm}

In the 1948 publication ``Fission Products of U$^{233}$'' Grummitt and Wilkinson reported the discovery of $^{123}$Sn \cite{1948Gru01}. Natural uranium and $^{233}$U targets were irradiated at Chalk River, Canada. ``After irradiation, a chemical separation of each element lying between arsenic and praseodymium was made. Thirty-one active isotopes were found and identified by their half-lives and their $\beta$-, and $\gamma$-ray absorption characteristics.''  In a table a half-life of 136~d was tentatively assigned to $^{123}$Sn. This half-life agrees with the currently accepted half-life of 129.2(4)~d. A 130~d half-life had been assigned to either $^{121}$Sn or $^{123}$Sn \cite{1946Lea01} and a 136~d half-life was assigned to Sn$^{>120}$ \cite{1946Gru01}. A 45(5)~m half-life, which could have corresponded to the 40.06(1)~m isomeric state, had been reported previously \cite{1936Liv02}. 47~m \cite{1937Poo01} and 40~m \cite{1939Liv01} half-lives were also observed in tin, however, no mass assignments were made. Finally, a 40~m half-life was assigned to Sn$^{<125}$ \cite{1947Ser01}.

\subsection{$^{124}$Sn}\vspace{0.0cm}

$^{124}$Sn was discovered by Aston in ``The Isotopes of Tin'' in 1922 \cite{1922Ast02}. The tin isotopes were identified with ``Half Tone'' plates installed at the Cavendish Laboratory mass spectrograph. ``Tin tetramethide was employed, and a group of eight lines corresponding approximately to atomic weights 116(c), 117(f), 118(b), 119(e), 120(a), 121(h), 122(g), 124(d) was definitely proved to be due to tin.'' The letters following the masses indicate the ordering of the observed intensity. The observation of the weakest isotope ($^{121}$Sn) proved to be incorrect.

\subsection{$^{125}$Sn}\vspace{0.0cm}

Livingood and Seaborg reported in 1939 the observation of $^{125}$Sn in the article ``New Periods of Radioactive Tin'' \cite{1939Liv01}. 5 MeV deuterons bombarded tin targets at the Berkeley Radiation Laboratory and radioactive decay curves were recorded. ``We have found in the tin precipitate, prepared by bombardment of tin with five-Mev deuterons, radioactivities with half-lives 9 minutes ($-$), 40 minutes ($-$), 26 hours ($-$), 10 days ($-$), about 70 days ($-$) and at least 400 days (sign unknown)... Inasmuch as neither we nor Pool, Cork and Thornton observe the 9-minute period when fast neutrons from Li+D are used, it is plausible to assign this period to Sn$^{125}$, obtained as the result of neutron capture by the heaviest stable isotope.'' This half-life agrees with the currently adopted value of 9.52(5)~m for an isomeric state. Half-lives of 8~m \cite{1936Nai01} and 6~m \cite{1936Nah01} had been reported for slow neutron activation of tin, however, no mass assignments were made.

\subsection{$^{126}$Sn}\vspace{0.0cm}

$^{126}$Sn was discovered in 1962 by Dropesky and Orth in ``A Summary of the Decay of Some Fission Product Tin and Antimony Isotopes'' \cite{1962Dro01}. $^{126}$Sn was produced by neutron induced fission at the Los Alamos Water Boiler Reactor and identified by $\gamma$-ray measurements following chemical separation. ``$^{126}$Sn, originally reported to have a half-life of 50 min, is now known to be very long-lived (T$_{1/2}$$\approx$10$^{5}$~years).'' This half-life agrees with the currently accepted half-life of 2.30(14)$\times$10$^{5}$~y. The quoted 50~min had been reported in 1951 \cite{1951Bar02} and could not been confirmed.

\subsection{$^{127}$Sn}\vspace{0.0cm}

In 1951, Barnes and Freedman published the article ``Some New Isotopes of Antimony and Tin'' which described the discovery of $^{127}$Sn \cite{1951Bar02}. $^{127}$Sn was produced at Los Alamos in neutron induced fission of $^{235}$U. Decay and absorption curves were measured following chemical separation.  ``From the amount of Sb activity obtained from the Sn as a function of time, the half-life of $^{127}$Sn was calculated; three experiments gave 83 min, 86 min, and 94 min, respectively.'' The extracted half-life of 1.5~h is close to the presently adopted value of 2.10(4)~h.

\subsection{$^{128}$Sn}\vspace{0.0cm}

In 1956 Fr\"anz et al. identified $^{128}$Sn in ``Die beiden Antimonisomere mit der Massenzahl 128'' \cite{1956Fra01}. Neutrons produced by bombarding beryllium with 28 MeV deuterons from the synchrocylotron at the CNEA, Buenos Aires, Argentina, induced fission of uranium. Gamma-rays were detected following chemical separation. ``Das Antimonisotop von 10,3 min Halbwertszeit l\"a\ss t sich von seiner Muttersubstanz, dem Spaltzinn von 57 min Halbwertszeit, sehr rein abtrennen... Man mu\ss\  daher der Isobarenreihe 57 min-Zinn $\rightarrow$ 10,3 min-Antimon die Massenzahl 128 zuordnen.'' (The 10.3~m antimon isotope can be easily separated from the mother substance tin with a half-life of 57~... Therefore the isobar chain: 57~m tin $\rightarrow$ 10.3~m antimon has to be assigned to mass 128.) The 57~m half-life agrees with the curently adopted value of 59.07(14)~m. The half-life had first been assigned to $^{130}$Sn \cite{1955Fra01}.

\subsection{$^{129}$Sn}\vspace{0.0cm}

Hagebo et al. reported the discovery of $^{129}$Sn in the 1962 article ``Radiochemical Studies of Isotopes of Antimony and Tin in the Mass Region 127-130'' \cite{1962Hag01}. 170 MeV protons bombarded uranium targets and $^{129}$Sn was identified by decay curve measurements following chemical separation. Although not explicitly mentioned, the proton irradiations were probably performed at the Institute of Nuclear Research in Amsterdam. ``Since the mass numbers of the antimony daughters are well known, mass assignments of the tin isotopes can be made unambiguously, i.e. both the 4.6~min and 2.2~hr have mass number 127, the 57~min is $^{128}$Sn, and both the 8.8~min and 1.0~hr have mass number 129.'' With the exception of the 1.0~h assignment to mass number 129, the reported half-lifes are in good agreement with currently accepted values. The half-life of 8.8(6)~m corresponds to the 6.9(1)~m isomeric state. A previously reported 1.8~h half-life \cite{1960Alv01} could not be confirmed.

\subsection{$^{130}$Sn}\vspace{0.0cm}

$^{130}$Sn was first correctly identified in 1972 by Izak and Amiel in ``Half-Lives and Gamma Rays of Tin Isotopes of Masses 129, 130, 131 and 132'' \cite{1972Iza01}. Neutrons from the Soreq IRR-1 reactor in Yavne, Israel, irradiated an enriched UO$_2$(NO$_3$)$_2$ solution. Decay curves, X-rays and $\gamma$-rays were recorded following chemical separation. ``A tin activity with a half-life of 3.69$\pm0.10$ min was assigned to $^{130}$Sn. The mean half-life obtained from twelve X-ray measurements was 3.72$\pm$0.18 min...'' The overall extracted half-life of 3.69(10)~m is included in the currently adopted average value of 3.72(7)~m. In 1956 Pappas and Wiles had reported a half-life of 2.6(3)~m \cite{1956Pap01} which could have corresponded to either the ground state (3.72(7)~m) or the 1.7(1)~m isomeric state. However, because none of the quoted half-lives for the other tin isotopes ($^{131}$Sn and $^{132}$Sn) were correct, we do not credit Pappas and Wiles with the discovery of $^{130}$Sn. The 57~m $^{128}$Sn half-life had initially been assigned to $^{130}$Sn \cite{1955Fra01}.

\subsection{$^{131,132}$Sn}\vspace{0.0cm}

In 1963 Greendale and Love reported the first observation of $^{131}$Sn and $^{132}$Sn in ``A Rapid Radiochemical Procedure for Tin'' \cite{1963Gre01}. The isotopes were produced by thermal-neutron induced fission of $^{235}$U and identified by chemical separation and decay curve measurements at the U.S. Naval Radiological Defense Laboratory. ``In the determination of the independent fission yield of a given tin isotope, the separated tin was allowed to decay to its known iodine descendent... It has been possible by the above technique to determine independent fission yields of the tin fission products in thermal neutron fission of uranium-235, tin half-lives, generic relationships, and also prominent gammaphotopeak energies from pulse-height distributions taken of the rapidly separated tin fractions.'' The measured half-live of 65(10)~s for $^{131}$Sn may correspond to either the ground state (56.0(5)~s) or the 58.4(5)~s isomeric state. The half-live for $^{132}$Sn (50(10)~s) is consistent with the currently adopted value 39.7(8)~s. Previously reported half-life values for $^{131}$Sn (3.4(5)~m \cite{1956Pap01} and 1.6~h \cite{1960Alv01}) and $^{132}$Sn (2.2~m \cite{1956Pap01}) could not be confirmed.

\subsection{$^{133}$Sn}\vspace{0.0cm}

In the 1973 article ``Proton Particle States in the Region Around $^{132}_{50}$Sn$_{82}$'' Borg et al. reported the observation of $^{133}$Sn \cite{1973Bor01}. Neutrons from the Studsvik R2-O swimming-pool reactor were used to fission $^{235}$U. $^{133}$Sn was separated and identified with the OSIRIS isotope separator on-line. In order to observe the short-lived $^{133}$Sn it was necessary to raise the energy threshold for the $\beta$-decay measurement: ``In this way a new half-life of 1.47$\pm$0.04~sec was observed, which is in reasonable agreement with the expected value for $^{133}$Sn.'' This half-life is included in the weighted average of the current value of 1.45(3)~s. Previously reported half-lives of 39(15)~s \cite{1963Gre01} and 55~s \cite{1966Str01} could not be confirmed.

\subsection{$^{134}$Sn}\vspace{0.0cm}

Shalev and Rudstam reported the discovery of $^{134}$Sn in ``Energy Spectra of Delayed Neutrons from Separated Fission Products'' in 1974 \cite{1974Sha01}. $^{134}$Sn was produced by neutron induced fission of $^{235}$U and identified with the OSIRIS isotope-separator on-line facility at Studsvik, Sweden. ''The precursor $^{134}$Sn was measured at the on-line position, using a tape speed of 0.4 cm/sec. The total number of fast neutrons (above 100 keV) was 4600. It was found that most of the neutron activity was associated with a half-life of about 1 sec, and hence we conclude that $^{134}$Sn is the dominant precursor and not $^{134}$Sb.'' This half-life is in agreement with the accepted half-life of 1.050(11)~s. A tentatively reported half-life of 20~s \cite{1963Gre01} could not been confirmed.

\subsection{$^{135-137}$Sn}\vspace{0.0cm}

Bernas {\it{et al.}} discovered $^{135}$Sn, $^{136}$Sn, and $^{137}$Sn in 1994 as reported in {\it{Projectile Fission at Relativistic Velocities: A Novel and Powerful Source of Neutron-Rich Isotopes Well Suited for In-Flight Isotopic Separation}} \cite{1994Ber01}. The isotopes were produced at the heavy-ion synchroton SIS at GSI using projectile fission of $^{238}$U at 750 MeV/nucleon on a lead target. ``Forward emitted fragments from $^{80}$Zn up to $^{155}$Ce were analyzed with the Fragment Separator (FRS) and unambiguously identified by their energy-loss and time-of-flight.'' The experiment yielded 193, 34, and 5 individual counts of $^{135}$Sn, $^{136}$Sn, and $^{137}$Sn, respectively.

\section{Discovery of $^{166-204}$Pt}

Thirty-nine platinum isotopes from A = $166-204$ have been discovered so far; these include 6 stable, 26 neutron-deficient and 7 neutron-rich isotopes. Many more additional neutron-rich nuclei are predicted to be stable with respect to neutron-emission and could be observed in the future. The mass surface towards the neutron dripline (the delineation where the neutron separation energy is zero) becomes very shallow. Thus the exact prediction of the location of the dripline is difficult and can vary substantially among the different mass models. As one example for a mass model we selected the HFB-14 model which is based on the Hartree-Fock-Bogoliubov method with Skyrme forces and a $\delta$-function pairing force \cite{2007Gor01}. According to this model $^{261}$Pt should be the last odd-even particle stable neutron-rich nucleus while the even-even particle stable neutron-rich nuclei should continue through $^{264}$Pt. Along the proton dripline four more isotopes ($^{162-165}$Pt) are predicted to be particle stable. In addition, it is estimated that 5 additional nuclei beyond the proton dripline could live long enough to be observed \cite{2004Tho01}. Thus about 68 isotopes have yet to be discovered corresponding to 64\% of all possible platinum isotopes.

In 2000, J.W. Arblaster published a review article entitled ``The Discoverers of the Platinum Isotopes'' \cite{2000Arb01}. Although he selected slightly different criteria for the discovery, our assignments agree in most of the cases. Since then only two additional isotopes ($^{203,204}$Pt) were discovered.

\begin{figure}
	\centering
	\includegraphics[scale=.5]{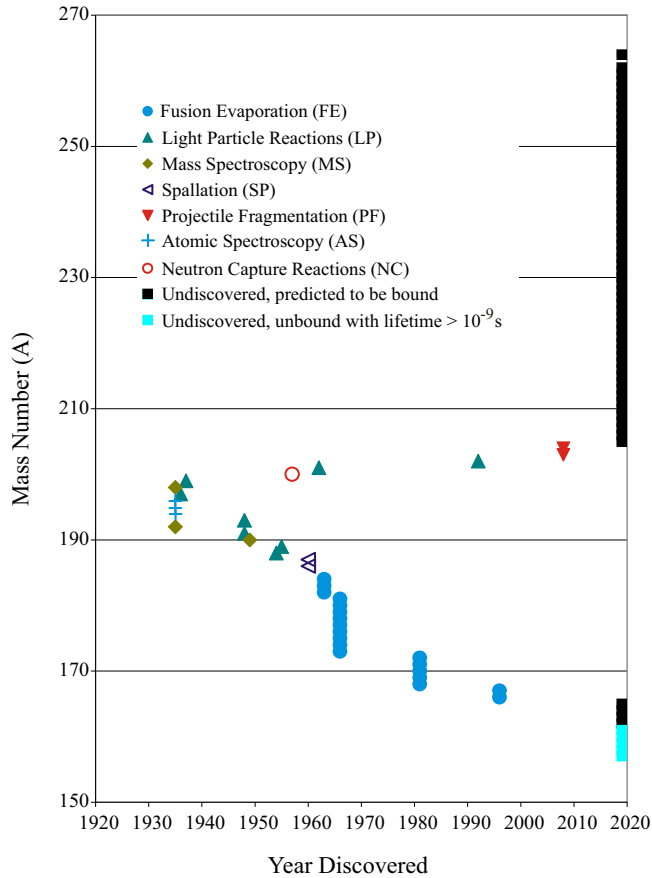}
	\caption{Platinum isotopes as a function of time when they were discovered. The different production methods are indicated. The solid black squares on the right hand side of the plot are isotopes predicted to be bound by the HFB-14 model. On the neutron-deficient side the light blue squares correspond to unbound isotopes predicted to have lifetimes larger than $\sim 10^{-9}$~s.}
\label{f:year-pt}
\end{figure}

Figure \ref{f:year-pt} summarizes the year of first discovery for all platinum isotopes identified by the method of discovery. The range of isotopes predicted to exist is indicated on the right side of the figure. The radioactive platinum isotopes were produced using heavy-ion fusion-evaporation reactions (FE), neutron capture reactions (NC), light-particle reactions (LP), spallation (SP), and projectile fragmentation of fission (PF). The stable isotopes were identified using mass spectroscopy (MS) or atomic spectroscopy (AS). Heavy ions are all nuclei with an atomic mass larger than A=4 \cite{1977Gru01}. Light particles also include neutrons produced by accelerators. In the following, the discovery of each platinum isotope is discussed in detail and a summary is presented in Table 1.

\subsection{$^{166-167}$Pt}\vspace{0.0cm}

$^{166-167}$Pt was discovered by Bingham et al. in 1996 and published in the paper ``Identification of $^{166}$Pt and $^{167}$Pt'' \cite{1996Bin01}. A $^{92}$Mo metal foil was bombarded by 357 and 384 MeV $^{78}$Kr beams at the ATLAS accelerator facility at Argonne National Laboratory. $^{166}$Pt and $^{167}$Pt were produced in fusion evaporation reactions and identified with the FMA fragment mass analyzer. ``These two figures demonstrate unambiguously the assignments of the 6832 and 6988 keV peaks to $^{168}$Pt and $^{167}$Pt, respectively. We deduced half-lives of 2.0(4) ms for $^{168}$Pt and 0.7(2) ms for the new isotope, $^{167}$Pt. ...we assign the previously unobserved 7110(15)-keV $\alpha$ peak to the new isotope $^{166}$Pt whose half-life was determined to be 0.3(1)~ms.'' The deduced half-lives of 0.3(1)~ms for $^{166}$Pt and 0.7(2)~ms for $^{167}$Pt are the currently accepted values.

\subsection{$^{168-171}$Pt}\vspace{0.0cm}

Hofmann et al. first identified $^{168}$Pt, $^{169}$Pt, $^{170}$Pt, and $^{171}$Pt in 1981. They published their results in ``New Neutron Deficient Isotopes in the Range of Elements Tm to Pt'' \cite{1981Hof01}. A $^{58}$Ni beam impinged on a tin target at the UNILAC linear accelerator. The $\alpha$-decay spectra of the evaporation residues were measured after the velocity filter SHIP. ``The lighter isotopes down to mass number 169 were identified in correlations to their well established daughters $^{167-165}$Os. The lightest isotope, $^{168}$Pt, could be identified by 4 correlated events to the daughter $^{164}$Os.'' The measured half-lives of 2.5$^{+2.5}_{-1.0}$~ms ($^{169}$Pt) and 6$^{+5}_{-2}$~ms for ($^{170}$Pt) are close to the currently accepted values of 7.0(2)~ms and 13.8(5)~ms, respectively. It should be mentioned that $^{171}$Pt was independently observed by Della Negra et al. \cite{1981Del01} and submitted less than two month after Hofmann et al.

\subsection{$^{172}$Pt}\vspace{0.0cm}

The discovery of $^{172}$Pt was reported in the 1981 article ``Alpha Decay Characteristics of Neutron Deficient Isotopes of Pt Isotopes Produced in $^{63}$Cu Induced Reactions on $^{112}$Sn and $^{113}$In Targets'' by Della Negra et al. \cite{1981Del01}. $^{112}$Sn and $^{113}$In targets were bombarded with 245-300 MeV $^{63}$Cu beams from the ALICE accelerator at Orsay. The evaporation residues were transported with a helium jet and deposited onto a metallic surface in front of an annular silicon detector. ``$^{173}$Pt is produced in this case by the reaction $^{112}$Sn($^{63}$Cu,pn) and the two other curves corresponding to (Cu,p2n) and (Cu,p3n) leading to $^{172}$Pt and $^{171}$Pt. The identification was confirmed by the study of [the] $^{113}$In(Cu,xn)Pt$^{176-x}$ excitation function for 3$\le$x$\le$5.'' The measured half-life of 120(10)~ms is close to the currently accepted half-life of 96(3)~ms.

\subsection{$^{173-181}$Pt}\vspace{0.0cm}

Siivola first observed $^{173-181}$Pt in 1966 and reported his results in ``Alpha-active Platinum Isotopes'' \cite{1966Sii01}. The Berkeley Heavy Ion Linear Accelerator HILAC was used to bombard $^{168,170,172}$Yb and $^{162,164}$Er targets with beams of $^{16}$O and $^{20}$Ne, respectively. The reaction products were deposited on an aluminum plate by helium gas flow. Alpha-particle decay was measured with a surface barrier counter and the isotopes were identified by excitation function measurements.  ``We conclude that the reaction observed in the $^{16}$O + Yb bombardments at 106 MeV excitation energy is ($^{16}$O,8n), and the others, with their maxima at 93 and 80 MeV, are ($^{16}$O,7n) and ($^{16}$O,6n), respectively. This and the regular behaviour of the Yb($^{16}$O,xn) reactions give unambigously the mass numbers down to $^{176}$Pt. The three lighter isotopes were assigned in a similar way using $^{20}$Ne + Er bombardments.'' The half-lives of 0.7(2)~s, 2.1(2)~s, 6.0(5)~s, 6.6(10)~s, 21.3(15)~s, 33(4)~s, 50(5)~s and 51(5)~s are in general agreement with the accepted values of 0.889(17)~s, 2.53(6)~s, 6.33(15)~s, 10.6(4)~s, 20.7(7)~s, 21.2(4)~s, 56(2)~s and 52.0(22)~s, respectively, for A = 174$-$181.

\subsection{$^{182-184}$Pt}\vspace{0.0cm}

$^{182}$Pt, $^{183}$Pt, and $^{184}$Pt were first observed by Graeffe in 1963. He reported his results in ``On the Alpha Activities of Platinum Isotopes'' \cite{1963Gra01}. An iridium target was bombarded by 50-150 MeV protons from the Gustav Werner Institute synchrocyclotron at Uppsala, Sweden. The $\alpha$-decay spectra were measured following chemical separation. ``The absence of an alpha activity due to Pt$^{184}$ is unlikely, so that the 20 min activity can be tentatively assigned to Pt$^{184}$... The hindered 6.5 min activity, whose alpha energy (4.74 MeV) exceeds that of the 20 min activity assigned to Pt$^{184}$ can be tentatively assigned to the next lighter odd isotope Pt$^{183}$, and the 2.5 min activity to the following even isotope Pt$^{182}$.'' The reported values of 2.5(5)~m, 6.5(10)~m and 20(2)~m agree with the currently accepted values of 3.0(2)~m, 6.5(10)~m and 17.3(2)~m, respectively, for A = 182$-$184. An earlier measurement of 2.5(5)~h \cite{1960Mal01} for the half-life of $^{184}$Pt could not be confirmed.

\subsection{$^{185}$Pt}\vspace{0.0cm}

$^{185}$Pt was first observed by Albouy et al. in 1960 with spallation reactions: ``Noveaux isotopes de p\'eriode courte obtenus par spallation de l'or'' \cite{1960Alb01}. The gold targets were bombared with 155 MeV protons from the Orsay synchro-cyclotron. ``L'intensit\'e obtenue pour les masses 187, 186 et 185 rend peu pr\'ecise la d\'etermination des \'energies des raies $\gamma$ correspondant aux masss 187 et 186 et ne nous a pas permis d'indentifier des raies $\gamma$ pour la cha\^ine de masses 185.'' (The intensity obtained for the masses 187, 186 and 185 makes a precise determination of the $\gamma$-ray energies corresponding to masses 187 and 186 possible but has not allowed us to identify $\gamma$-rays for the mass of 185.) The quoted half-life of 1.2~h agrees with the accepted value of 70.9(2.4)~m.

\subsection{$^{186,187}$Pt}\vspace{0.0cm}

Baranov et al. reported the first identification of $^{186}$Pt and $^{187}$Pt in the 1961 article ``New Iridium and Platinum Isotopes: Ir$^{184}$ and Pt$^{187}$'' \cite{1961Bar01}. The isotopes were produced by bombarding a gold target with 660 MeV protons from the Dubna Joint Institute for Nuclear Research synchrocyclotron and identified following chemical separation. ``The period determined in this manner for Pt$^{187}$ was 2.0$\pm$0.4 hours. For control purposes we also determined the period of the known platinum isotope Pt$^{186}$, from the intensity of the L-137 line belonging to Ir$^{186}$. We obtained for Pt$^{186}$ a period of 2.5$\pm$0.3 hours, which is in good agreement with the data of Smith \& Hollander.'' Smith \& Hollander \cite{1955Smi01} had assigned a half-life of 2.5~h to $^{187}$Pt based on the $\gamma$-ray spectra of the daughter $^{187}$Ir. However, the assignment of the observed $\gamma$-rays was later changed to $^{186}$Ir \cite{1958Dia01}. Similar half-lives were reported by some of the same authors in the same issue \cite{1960Mal01}. About 6 month later the observation of $^{186}$Pt and $^{187}$Pt was independently reported by Albouy et al. \cite{1960Alb01}.
The half-lives of 2.5(3)~h ($^{186}$Pt) and 2.0(4)~h ($^{187}$Pt) agree with the presently accepted values of 2.08(5)~h and 2.35(3)~h, respectively.

\subsection{$^{188}$Pt}\vspace{0.0cm}

Naumann was the first to observe $^{188}$Pt and reported his results in the 1954 paper ``Identification of Platinum-188'' \cite{1954Nau01}. 50-MeV protons from the Nevis and Harvard synchrocyclotrons bombarded a metallic iridium foil. Decay curves were measured with an Amperex 200C Geiger-M\"uller counter. ``The reappearance of the 10-day decay component preceded by the short period growth suggests that this half-life be assigned to Pt$^{188}$.'' The reported half-life of 10.3(4)~d agrees with the currently accepted value of 10.2(3)~d.

\subsection{$^{189}$Pt}\vspace{0.0cm}

The discovery of $^{189}$Pt was reported in ``Radiochemical Study of Neutron-Deficient Chains in the Noble Metal Region'' by Smith and Hollander in 1955 \cite{1955Smi01}. A set of stacked iridium and aluminum foils were bombarded with 32-MeV protons from the Berkeley proton linear accelerator. Decay curves of the chemical separated reaction products were recorded with a Geiger counter. ``An $\sim$12-hour activity in platinum was first observed in 1950 by Thompson and Rasmussen from 50-Mev proton bombardments of iridium, but a mass assignment was not made at that time. With the aid of J.O. Rasmussen, this activity has now been assigned to Pt$^{189}$ by means of proton excitation function experiments in which its yield from iridium is compared with that of 3.0-day Pt$^{191}$ produced from the (p,3n) reaction on Ir$^{193}$.'' The reported half life of 10.5(10)~h is consistent with the currently accepted value of 10.87(12)~h. The reference mentioned in the quote was unpublished \cite{1950Tho04}.

\subsection{$^{190}$Pt}\vspace{0.0cm}

$^{190}$Pt was first reported by Duckworth et al. in ``A New Stable Isotope of Platinum'' \cite{1949Duc01} in 1949. Platinum isotopes were produced in an ion source and detected and analyzed by a double-focusing mass spectrograph. ``With a spark between a platinum-iridium and a copper electrode a faint line appeared at mass 190 after an exposure of three hours. The electrodes were replaced by pure platinum electrodes. With an hour's exposure the faint 190 appeared. Similarly with two platinum electrodes of commercial purity the line at 190 was clearly visible after an hour's exposure.''

\subsection{$^{191}$Pt}\vspace{0.0cm}

In 1947 Wilkinson described the first observation of $^{191}$Pt in ``Some Isotopes of Platinum and Gold'' \cite{1948Wil01}. Platinum and iridium targets were bombarded with $\alpha$-particles, deuterons and neutrons from the 60-inch Crocker Laboratory cyclotron. Decay curves were measured following chemical separation. ``{\it 3.0-day platinum.}$-$Previously unreported, this isotope which is distinguished from the 4.3-day isotope by its 0.5-Mev electron and strong 0.57-Mev $\gamma$-ray has been observed in low yield in Pt+d, Pt+fast neutron, and Ir+$\alpha$ bombardments... The activity has been assigned provisionally to mass 191.'' The measured half-life of 3.00(2)~d is consistent with the currently accepted value of 2.83(2)~d.

\subsection{$^{192}$Pt}\vspace{0.0cm}

Dempster identified $^{192}$Pt for the first time in the 1935 article ``Isotopic Constitution of Platinum and Rhodium'' \cite{1935Dem01}. An alloy of platinum with 10\% rhodium were used as electrodes of a spark for the source of a spectrograph. ``The analysis of the platinum ions from a high-frequency spark, using a new spectrograph, shows that this element consists of five isotopes with masses 192, 194, 195, 196, 198.''

\subsection{$^{193}$Pt}\vspace{0.0cm}

In 1947 Wilkinson described the first observation of $^{193}$Pt in ``Some Isotopes of Platinum and Gold'' \cite{1948Wil01}. The 60-inch Crocker Laboratory cyclotron bombarded platinum and iridium targets with $\alpha$-particles, deuterons and neutrons. Decay curves were measured following chemical separation. ``...The activity is attributed to Pt$^{193}$ decaying by orbital electron capture for the following reasons. The isotope is formed in the deuteron, fast and thermal neutron bombardments of platinum, and also the deuteron and $\alpha$-particle bombardment of iridium in yields agreeing with allocation to mass 193.'' The reported half-life of 4.33(3)~d corresponds to an isomeric state. A previously measured half-life of 49~m \cite{1936Cor01} could not be confirmed.

\subsection{$^{194-196}$Pt}\vspace{0.0cm}

$^{194}$Pt, $^{195}$Pt and $^{196}$Pt were first identified by Fuchs and Kopfermann in the 1935 article ``\"Uber die Isotopen des Platins'' \cite{1935Fuc01}. The masses of these isotopes were determined by measuring the isotope shift of the visible platinum lines. ``In der Erwartung, da\ss\ im Spektrum des Platins eine an schweren Elementen h\"aufig beobachtete Isotopieverschiebung auftreten w\"urde, haben wir, um das Isotopenproblem dieses Elementes zu l\"osen, eine Hyperfeinstrukturanalyse der im Sichtbaren gelegenen PtI-Linien durchgef\"uhrt... Der Vergleich mit dem chemischen Atomgewicht (195.2) l\"a\ss t nur die Deutung zu, da\ss\ es sich bei den drei gefundenen Isotopen um die Pt-Isotopen mit den Massenzahlen 194,195 und 196 handelt, deren Mischungsverh\"ahltnis auf Grund unserer Intensit\"atssch\"atzungen ungef\"ahr 5:8:8 betr\"agt.'' [With the expectation that the spectra of platinum will exhibit an isotope shift common to many of the heavy elements, we conducted a hyperfine structure analysis of the visible PtI lines in order to solve the isotope problem of this element... The only interpretation from the comparison with the chemical atomic weight (195.2) is that the three observed isotopes correspond to the Pt-isotopes with the mass numbers 194, 195, and 196; according to our intensity estimates the abundance ratio is approximately 5:8:8.] The mass spectroscopic identification of these isotopes were submitted less than two weeks later \cite{1935Dem01}.

\subsection{$^{197}$Pt}\vspace{0.0cm}

Cork and Lawrence reported the discovery of $^{197}$Pt in the 1936 publication ``The Transmutation of Platinum by Deuterons'' \cite{1936Cor01}. Deuterons accelerated to 5 MV by a magnetic resonance accelerator bombarded a stack of platinum foils. The resulting isotopes were separated by chemical means and the decay curves of the individual foils were recorded. ``Because of the greater abundance of Pt$^{196}$ the 14.5-hr. electron activity of platinum can be reasonably ascribed to Pt$^{197}$, which decays to gold...'' The reported half-life of 14.5~hr is close to the currently accepted value of 19.8915(19)~h.

\subsection{$^{198}$Pt}\vspace{0.0cm}

Dempster identified $^{198}$Pt for the first time in the 1935 article ``Isotopic Constitution of Platinum and Rhodium'' \cite{1935Dem01}. An alloy of platinum with 10\% rhodium were used as electrodes of a spark for the source of a spectrograph. ``The analysis of the platinum ions from a high-frequency spark, using a new spectrograph, shows that this element consists of five isotopes with masses 192, 194, 195, 196, 198.''

\subsection{$^{199}$Pt}\vspace{0.0cm}

McMillan et al. observed $^{199}$Pt in the 1937 article ``Neutron-Induced Radioactivity of the Noble Metals'' \cite{1937McM01}. Slow neutrons irradiated platinum targets which were subsequently chemical separated and their activity was measured. ``Reference to the isotope chart show that one would expect Pt$^{199}$ to form unstable Au$^{199}$. We made successive separations of gold from activated platinum to find which platinum period is its parent, and found that it does not come from the 18-hr. period, but most probably does come from the 31-min. period.'' The reported half life of 31~min is in agreement with the currently accepted value of 30.80(21)~min. Two activities of 50~min \cite{1935Ama01} and 36~min \cite{1935McL01} had previously been reported for platinum without mass assignment. Also, a half-life of 49~min had been assigned to $^{193}$Pt \cite{1936Cor01}.

\subsection{$^{200}$Pt}\vspace{0.0cm}

Roy et al. described the discovery of $^{200}$Pt in ``New Radioisotope of Platinum$\--$Pt$^{200}$'' \cite{1957Roy02}. $^{198}$Pt targets were irradiated by neutrons from the Chalk River NRX reactor to produce $^{200}$Pt by successive neutron capture. The presence of $^{200}$Pt was determined by milking the $^{200}$Au daughter and measuring the activities. ``Parent-daughter isolation experiments were performed to establish both the half-life of Pt$^{200}$ and its genetic relationship to Au$^{200}$... The average value for the half-life of Pt$^{200}$ is 11.5$\pm$1.0~hr.'' This half-life agrees with the currently accepted value of 12.6(3)~h.

\subsection{$^{201}$Pt}\vspace{0.0cm}

$^{201}$Pt was first reported by Facetti et al. in the 1962 report ``A New Isotope, Pt$^{201}$'' \cite{1962Fac01}. Neutrons from the Puerto Rico Nuclear Center nuclear reactor were used to irradiate mercury compounds. The activated samples were chemically separated and their decay was measured with a GM counter. ``From all these facts, it follows that the observed half-life of 2.3$\pm$0.2~min is due to the disintegration of a new isotope: Pt$^{201}$.'' The reported half-life of 2.3(2)~m is in good agreement with the currently accepted value of 2.5(1)~m.

\subsection{$^{202}$Pt}\vspace{0.0cm}

The first observation of $^{202}$Pt was described in 1992 by Shi et al. in ``Identification of a New Neutron Rich Isotope $^{202}$Pt'' \cite{1992Shi01}. 250 MeV protons from the Shanghai Institute of Nuclear Research K=40 cyclotron bombarded a beryllium target to produce neutrons, which then irradiated a mercury target. $^{202}$Pt was identified by measuring the $\gamma$-ray spectra of the chemically separated activities with a HpGe detector. ``The gamma ray 439.6 keV from the decay $^{202}$Pt $\to ^{202}$Au $\to ^{202}$Hg was detected... The half-life of $^{202}$Pt is determined to be 43.6$\pm$15.2~h.'' This half-life is the currently accepted value.

\subsection{$^{203,204}$Pt}\vspace{0.0cm}

The first refereed publication of the observation of $^{203}$Pt and $^{204}$Pt was the 2008 paper ``Single-particle Behavior at N = 126: Isomeric Decays in Neutron-rich $^{204}$Pt'' by Steer et al. \cite{2008Ste01}. A 1 GeV/A $^{208}$Pb beam from the SIS-18 accelerator at GSI impinged on a $^9$Be target and the projectile fragments were selected and identified in flight by the FRagment Separator FRS. The observation of $^{203}$Pt is not specifically mentioned but $^{203}$Pt events are clearly visible and identified in the particle identification plot in the first figure. Details of the $^{203}$Pt had previously been published by the same group in two different conference proceedings \cite{2005Kur01,2007Ste01}.``The results for $^{204}$Pt were obtained from four different magnetic rigidity settings of the FRS. A total of 9.3 $\times 10^{4}$ $^{204}$Pt ions was implanted in the stopper.'' The $^{204}$Pt data also had been presented previously in two conference proceedings \cite{2007Ste01,2007Pod01}.

\section{Summary}
The discoveries of the known calcium, indium, tin, and platinum isotopes have been compiled and the methods of their production discussed.

The discovery of most of the calcium isotopes was straight forward. Only two isotopes ($^{38}$Ca and $^{39}$Ca) were initially identified incorrectly. $^{37}$Ca is one of the rare cases where two papers reporting the discovery were submitted on the same day.

The first measured half-lives of several indium isotopes ($^{104}$In, $^{108}$In, $^{111}$In, $^{112}$In, $^{114}$In, and $^{121}$In) were incorrect. The half-lives of $^{114}$In and $^{116}$In were first reported without a definite mass assignment. The discovery of $^{105}$In was only published two years after submission of the article during which time $^{105}$In was reported by other authors. The discovery of $^{133}$In appeared in a refereed journal 15 years after it had been reported in a conference proceeding.

The correct identification of many of the tin isotopes proved difficult. $^{121}$Sn was first incorrectly reported as a stable isotope. The half-life measurements of $^{107,126,129,131-134}$Sn were initially incorrect. The half-lives of $^{110}$Sn and $^{128}$Sn were first incorrectly assigned to $^{108}$Sn and $^{130}$Sn, respectively. In addition, the half-lives of $^{121}$Sn, $^{123}$Sn, and $^{125}$Sn were initially measured without a firm mass identification. Finally, the half-life of $^{110}$Sn had been reported in an unpublished thesis ten years prior to publication in a refereed journal.

Most of the assignments for the platinum isotopes agree with the review article by Arblaster \cite{2000Arb01}. The exceptions are: (1) $^{172}$Pt where Arblaster credits unpublished work by Cabot as quoted in an overview paper by Gauvin et al. \cite{1975Gau01}; (2) $^{186}$Pt which is credited to Albouy et al. \cite{1960Alb01}, however the identification by Baranov et al. \cite{1961Bar01} published a few months earlier is correct; and (3) the stable isotopes $^{194-196}$Pt where we found a paper by Fuchs and Kopfermann \cite{1935Fuc01} which was submitted less than two weeks prior to the mass spectroscopic work by Dempster \cite{1935Dem01}. In addition, the half-lives for $^{184}$Pt and $^{193}$Pt were first reported incorrectly and $^{186}$Pt was first identified as $^{187}$Pt. The half-life of $^{199}$Pt was initially measured without mass identification. Finally, $^{203}$Pt has yet to be specifically described in the refereed literature.

\ack

The main research on the individual elements were performed by SA (indium and tin) and JLG (calcium and platinum). This work was supported by the National Science Foundation under grant No. PHY06-06007 (NSCL).

%%% Here we use thebibliography environment to produce the reference list,
%%% but you can use BibTeX as well:
\bibliography{../isotope-discovery-references}

\newpage

%%% Please start a new page by uncommenting the next
\newpage

\TableExplanation

\bigskip
\renewcommand{\arraystretch}{1.0}

\section{Table 1.\label{tbl1te} Discovery of calcium, indium, tin, and platinum isotopes }
\begin{tabular*}{0.95\textwidth}{@{}@{\extracolsep{\fill}}lp{5.5in}@{}}
\multicolumn{2}{p{0.95\textwidth}}{ }\\

Isotope & Calcium, indium, tin, and platinum isotope \\
Author & First author of refereed publication \\
Journal & Journal of publication \\
Ref. & Reference \\
Method & Production method used in the discovery: \\
   & FE: fusion evaporation \\
    & LP: light-particle reactions (including neutrons) \\
    & MS: mass spectroscopy \\
    & AS: atomic spectroscopy \\
    & PN: photo nuclear reactions \\
    & NC: neutron-capture reactions \\   & NF: neutron induced fission \\
    & CPF: charged particle induced fission \\
    & SP: spallation reactions \\
    & PF: projectile fragmentation or fission \\
Laboratory & Laboratory where the experiment was performed\\
Country & Country of laboratory\\
Year & Year of discovery \\
\end{tabular*}
\label{tableI}

\datatables % This command is necessary to get the table names in toc

%% One-page data tables are also best formatted using the longtable
%% environment:
%\begin{longtable}{c}
%\caption{This is the First Data Table}\\
%\endhead\\
%\end{longtable}

%% If the table is to span over the whole text width, we set:

\setlength{\LTleft}{0pt}
\setlength{\LTright}{0pt}

% To avoid ``Overfull \hboxes...'' decrease the intercolumn spacing:

\setlength{\tabcolsep}{0.5\tabcolsep}

\renewcommand{\arraystretch}{1.0}

\footnotesize % we need to squeeze the font size a lot!

\begin{longtable}{@{\extracolsep\fill}llllllll@{}}
\caption{Discovery of Calcium, Indium, Tin, and Platinum Isotopes. See page\ \pageref{tbl1te} for Explanation of Tables}
Isotope & Author & Journal & Ref. & Method & Laboratory & Country & Year\\
\hline\\
\endfirsthead\\
\caption[]{(continued)}
Isotope & Author & Journal & Ref. & Method & Laboratory & Country & Year\\
\hline\\
\endhead
$^{35}$Ca & J. Aysto & Phys. Rev. Lett. &\cite{1985Ays01}& LP & Berkeley & USA &1985 \\
$^{36}$Ca & R.E. Tribble & Phys. Rev. C &\cite{1977Tri01}& LP & Texas A\&M& USA &1977 \\
$^{37}$Ca & J.C. Hardy & Phys. Rev. Lett. &\cite{1964Har01}& LP & McGill & Canada &1964 \\
 & P. L. Reeder & Phys. Rev. Lett. &\cite{1964Ree01}& LP & Brookhaven & USA &1964 \\
$^{38}$Ca & J.C. Hardy & Phys. Lett. &\cite{1966Har01}& LP & Oxford & UK &1966 \\
$^{39}$Ca & O. Huber & Helv. Phys. Acta &\cite{1943Hub01}& PN & Zurich & Switzerland &1943 \\
$^{40}$Ca & A.J. Dempster & Phys. Rev. &\cite{1922Dem01}& MS & Chicago & USA &1922 \\
$^{41}$Ca & W.L. Davidson & Phys. Rev. &\cite{1939Dav01}& LP & Yale & USA &1939 \\
$^{42}$Ca & F.W. Aston & Nature &\cite{1934Ast02}& MS & Cambridge & UK &1934 \\
$^{43}$Ca & F.W. Aston & Nature &\cite{1934Ast02}& MS & Cambridge & UK &1934 \\
$^{44}$Ca & A.J. Dempster & Phys. Rev. &\cite{1922Dem01}& MS & Chicago & USA &1922 \\
$^{45}$Ca & H. Walke & Phys. Rev. &\cite{1940Wal02}& LP & Berkeley & USA &1940 \\
$^{46}$Ca & A.O. Nier & Phys. Rev. &\cite{1938Nie01}& MS & Harvard & USA &1938 \\
$^{47}$Ca & R.E. Batzel & Phys. Rev. &\cite{1951Bat01}& SP & Berkeley & USA &1951 \\
$^{48}$Ca & A.O. Nier & Phys. Rev. &\cite{1938Nie01}& MS & Harvard & USA &1938 \\
$^{49}$Ca & E. der Mateosian & Phys. Rev. &\cite{1950Mat01}& NC & Argonne & USA &1950 \\
$^{50}$Ca & Y. Shida & Phys. Lett. &\cite{1964Shi01}& LP & Kawasaki & Japan &1964 \\
$^{51}$Ca & A. Huck & Phys. Rev. C &\cite{1980Huc01}& SP & CERN & Switzerland &1980 \\
$^{52}$Ca & A. Huck & Phys. Rev. C &\cite{1985Huc01}& SP & CERN & Switzerland &1985 \\
$^{53}$Ca & M. Langevin & Phys. Lett. B &\cite{1983Lan01}& SP & CERN & Switzerland &1983 \\
$^{54}$Ca & M. Bernas & Phys. Lett. B &\cite{1997Ber01}& PF & Darmstadt & Germany &1997 \\
$^{55}$Ca & M. Bernas & Phys. Lett. B &\cite{1997Ber01}& PF & Darmstadt & Germany &1997 \\
$^{56}$Ca & M. Bernas & Phys. Lett. B &\cite{1997Ber01}& PF & Darmstadt & Germany &1997 \\
$^{57}$Ca & O.B. Tarasov & Phys. Rev. Lett. &\cite{2009Tar01}& PF & Michigan State & USA &2009 \\
$^{58}$Ca & O.B. Tarasov & Phys. Rev. Lett. &\cite{2009Tar01}& PF & Michigan State & USA &2009 \\
&  &  & &  &  & & \\
&  &  & &  &  & & \\
$^{98}$In & R. Schneider & Z. Phys. A &\cite{1994Sch01}& PF & Darmstadt & Germany &1994 \\
$^{99}$In & R. Schneider & Z. Phys. A &\cite{1994Sch01}& PF & Darmstadt & Germany &1994 \\
$^{100}$In & W. Kurcewicz & Z. Phys. A &\cite{1982Kur01}& FE & Darmstadt & Germany &1982 \\
$^{101}$In & M. Huyse & Z. Phys. A &\cite{1988Huy01}& FE & Louvain-la-Neuve & Belgium &1988 \\
$^{102}$In & B. Beraud & Z. Phys. A &\cite{1981Ber01}& FE & Grenoble & France &1981 \\
$^{103}$In & G. Lhersonneau & Phys. Rev. C &\cite{1978Lhe01}& FE & Louvain-la-Neuve & Belgium &1978 \\
$^{104}$In & B.J. Varley & J. Phys. G &\cite{1977Var01}& FE & Manchester & UK &1977 \\
$^{105}$In & J. Rivier & Radiochim. Acta &\cite{1975Riv01}& LP & Grenoble & France &1975 \\
$^{106}$In & R.C. Catura & Phys. Rev. &\cite{1962Cat01}& LP & UCLA & USA &1962 \\
$^{107}$In & E.C. Mallary & Phys. Rev. &\cite{1949Mal01}& LP & Ohio State & USA &1949 \\
$^{108}$In & E.C. Mallary & Phys. Rev. &\cite{1949Mal01}& LP & Ohio State & USA &1949 \\
$^{109}$In & S.N. Goshal & Phys. Rev. &\cite{1948Gho01}& LP & Berkeley & USA &1948 \\
$^{110}$In & S.W. Barnes & Phys. Rev. &\cite{1939Bar01}& LP & Rochester & USA &1939 \\
$^{111}$In & D.J. Tendam & Phys. Rev. &\cite{1947Ten01}& LP & Purdue & USA &1947 \\
$^{112}$In & D.J. Tendam & Phys. Rev. &\cite{1947Ten01}& LP & Purdue & USA &1947 \\
$^{113}$In & M. Wehrli & Naturwiss. &\cite{1934Weh01}& MS & Basel & Switzerland &1934 \\
$^{114}$In & J.L. Lawson & Phys. Rev. &\cite{1937Law01}& LP & Michigan & USA &1937 \\
$^{115}$In & F.W. Aston & Nature &\cite{1924Ast01}& MS & Cambridge & UK &1924 \\
$^{116}$In & J.L. Lawson & Phys. Rev. &\cite{1937Law01}& NC & Michigan & USA &1937 \\
$^{117}$In & J.M. Cork & Phys. Rev. &\cite{1937Cor01}& LP & Michigan & USA &1937 \\
$^{118}$In & R.B. Duffield & Phys. Rev. &\cite{1949Duf02}& PN & Illinois & USA &1949 \\
$^{119}$In & R.B. Duffield & Phys. Rev. &\cite{1949Duf02}& PN & Illinois & USA &1949 \\
$^{120}$In & C.L. McGinnis& Phys. Rev. &\cite{1958McG01}& LP & Nat. Bureau of Standards  & USA &1958 \\
$^{121}$In & H. Yuta & Nucl. Phys. &\cite{1960Yut01}& PN & Tohoku & Japan &1960 \\
$^{122}$In & J. Kantele & Phys. Rev. &\cite{1963Kan01}& LP & Arkansas & USA &1963 \\
$^{123}$In & H. Yuta & Nucl. Phys. &\cite{1960Yut01}& PN & Tohoku & Japan &1960 \\
$^{124}$In & M. Karras & Phys. Rev. &\cite{1964Kar01}& LP & Arkansas & USA &1964 \\
$^{125}$In & K. Fritze & Radiochim. Acta &\cite{1967Fri01}& NF & McMaster & Canada &1967 \\
$^{126}$In & B. Grapengiesser & J. Inorg. Nucl. Chem. &\cite{1974Gra01}& NF & Studsvik & Sweden &1974 \\
$^{127}$In & K. Aleklett & Nucl. Phys. A &\cite{1975Ale01}& NF & Studsvik & Sweden &1975 \\
$^{128}$In & K. Aleklett & Nucl. Phys. A &\cite{1975Ale01}& NF & Studsvik & Sweden &1975 \\
$^{129}$In & K. Aleklett & Nucl. Phys. A &\cite{1975Ale01}& NF & Studsvik & Sweden &1975 \\
$^{130}$In & A. Kerek & Nucl. Phys. A &\cite{1973Ker01}& NF & Studsvik & Sweden &1973 \\
$^{131}$In & E. Lund & Phys. Rev. C &\cite{1976Lun01}& NF & Studsvik & Sweden &1976 \\
$^{132}$In & A. Kerek & Phys. Lett. B &\cite{1973Ker02}& NF & Studsvik & Sweden &1973 \\
$^{133}$In & P. Hoff & Phys. Rev. Lett. &\cite{1996Hof01}& SP & CERN & Switzerland &1996 \\
$^{134}$In & P. Hoff & Phys. Rev. Lett. &\cite{1996Hof01}& SP & CERN & Switzerland &1996 \\
$^{135}$In & I. Dillmann & Eur. Phys. J. A &\cite{2002Dil01}& SP & CERN & Switzerland &2002 \\
&  &  & &  &  & & \\
&  &  & &  &  & & \\
$^{100}$Sn & R. Schneider & Z. Phys. A &\cite{1994Sch01}& PF & Darmstadt & Germany &1994 \\
$^{101}$Sn & R. Schneider & Z. Phys. A &\cite{1994Sch01}& PF & Darmstadt & Germany &1994 \\
$^{102}$Sn & M. Lewitowicz & Phys. Lett. B &\cite{1994Lew01}& PF & GANIL & France &1994 \\
$^{103}$Sn & P. Tidemand-Petersson & Z. Phys. A &\cite{1981Tid01}& FE & Darmstadt & Germany &1981 \\
$^{104}$Sn & K. Deneffe & J. Phys. G &\cite{1985Den01}& FE & Louvain-la-Neuve & Belgium &1985 \\
$^{105}$Sn & P. Tidemand-Petersson & Z. Phys. A &\cite{1981Tid01}& FE & Darmstadt & Germany &1981 \\
$^{106}$Sn & V.N. Burminskii & JETP Lett. &\cite{1975Bur01}& LP & Almaty & Kaszakhstan &1975 \\
$^{107}$Sn & H.C. Hseuh & Phys. Rev. C &\cite{1976Hse01}& LP & St. Louis & USA &1976 \\
$^{108}$Sn & T. Yamazaki & Phys. Rev. Lett. &\cite{1968Yam01}& LP & Berkeley & USA &1968 \\
$^{109}$Sn & D.E. Khulelidze & Bull. Acad. Sci. USSR &\cite{1966Khu01}& LP & Dubna & Russia &1966 \\
$^{110}$Sn & G. Bassani & Phys. Rev. &\cite{1965Bas01}& LP & Minnesota & USA &1965 \\
$^{111}$Sn & R.A. Hinshaw & Phys. Rev. &\cite{1949Hin01}& LP & Ohio State & USA &1949 \\
$^{112}$Sn & F.W. Aston & Nature &\cite{1927Ast02}& MS & Cambridge & UK &1927 \\
$^{113}$Sn & J.J. Livingood & Phys. Rev. &\cite{1939Liv01}& LP & Berkeley & USA &1939 \\
$^{114}$Sn & F.W. Aston & Nature &\cite{1927Ast02}& MS & Cambridge & UK &1927 \\
$^{115}$Sn & F.W. Aston & Nature &\cite{1927Ast02}& MS & Cambridge & UK &1927 \\
$^{116}$Sn & F.W. Aston & Nature &\cite{1922Ast02}& MS & Cambridge & UK &1922 \\
$^{117}$Sn & F.W. Aston & Nature &\cite{1922Ast02}& MS & Cambridge & UK &1923 \\
$^{118}$Sn & F.W. Aston & Nature &\cite{1922Ast02}& MS & Cambridge & UK &1924 \\
$^{119}$Sn & F.W. Aston & Nature &\cite{1922Ast02}& MS & Cambridge & UK &1925 \\
$^{120}$Sn & F.W. Aston & Nature &\cite{1922Ast02}& MS & Cambridge & UK &1926 \\
$^{121}$Sn & M. Lindner & Phys. Rev. &\cite{1948Lin01}& LP & Berkeley & USA &1948 \\
$^{122}$Sn & F.W. Aston & Nature &\cite{1922Ast02}& MS & Cambridge & UK &1928 \\
$^{123}$Sn & W.E. Grummitt & Nature &\cite{1948Gru01}& NF & Chalk River & Canada &1948 \\
$^{124}$Sn & F.W. Aston & Nature &\cite{1922Ast02}& MS & Cambridge & UK &1922 \\
$^{125}$Sn & J.J. Livingood & Phys. Rev. &\cite{1939Liv01}& LP & Berkeley & USA &1939 \\
$^{126}$Sn & B.J. Dropesky & J. Inorg. Nucl. Chem. &\cite{1962Dro01}& NF & Los Alamos & USA &1962 \\
$^{127}$Sn & J.W. Barnes & Phys. Rev. &\cite{1951Bar02}& NF & Los Alamos & USA &1951 \\
$^{128}$Sn & I. Franz & Z. Naturforsch. &\cite{1956Fra01}& NF & Buenos Aires & Argentina &1956 \\
$^{129}$Sn & E. Hagebo& J. Inorg. Nucl. Chem. &\cite{1962Hag01}& CPF & Amsterdam & Netherlands &1962 \\
$^{130}$Sn & T. Izak & J. Inorg. Nucl. Chem. &\cite{1972Iza01}& NF & Soreq & Israel &1972 \\
$^{131}$Sn & A.E. Greendale & Anal. Chem. &\cite{1963Gre01}& NF & Naval Radiological Defense Laboratory & USA &1963 \\
$^{132}$Sn & A.E. Greendale & Anal. Chem. &\cite{1963Gre01}& NF & Naval Radiological Defense Laboratory & USA &1963 \\
$^{133}$Sn & S. Borg & Nucl. Phys. A &\cite{1973Bor01}& NF & Studsvik & Sweden &1973 \\
$^{134}$Sn & S. Shalev & Nucl. Phys. A &\cite{1974Sha01}& NF & Studsvik & Sweden &1974 \\
$^{135}$Sn & M. Bernas & Phys. Lett. B &\cite{1994Ber01}& PF & Darmstadt & Germany &1994 \\
$^{136}$Sn & M. Bernas & Phys. Lett. B &\cite{1994Ber01}& PF & Darmstadt & Germany &1994 \\
$^{137}$Sn & M. Bernas & Phys. Lett. B &\cite{1994Ber01}& PF & Darmstadt & Germany &1994 \\
&  &  & &  &  & & \\
&  &  & &  &  & & \\
$^{166}$Pt & C.R. Bingham & Phys. Rev. C &\cite{1996Bin01}& FE & Argonne & USA &1996 \\
$^{167}$Pt & C.R. Bingham & Phys. Rev. C &\cite{1996Bin01}& FE & Argonne & USA &1996 \\
$^{168}$Pt & S. Hofmann & Z. Phys. A &\cite{1981Hof01}& FE & Darmstadt & Germany &1981 \\
$^{169}$Pt & S. Hofmann & Z. Phys. A &\cite{1981Hof01}& FE & Darmstadt & Germany &1981 \\
$^{170}$Pt & S. Hofmann & Z. Phys. A &\cite{1981Hof01}& FE & Darmstadt & Germany &1981 \\
$^{171}$Pt & S. Hofmann & Z. Phys. A &\cite{1981Hof01}& FE & Darmstadt & Germany &1981 \\
$^{172}$Pt & S. Della Negra & Z. Phys. A &\cite{1981Del01}& FE & Orsay & France &1981 \\
$^{173}$Pt & A. Siivola & Nucl. Phys. &\cite{1966Sii01}& FE & Berkeley & USA &1966 \\
$^{174}$Pt & A. Siivola & Nucl. Phys. &\cite{1966Sii01}& FE & Berkeley & USA &1966 \\
$^{175}$Pt & A. Siivola & Nucl. Phys. &\cite{1966Sii01}& FE & Berkeley & USA &1966 \\
$^{176}$Pt & A. Siivola & Nucl. Phys. &\cite{1966Sii01}& FE & Berkeley & USA &1966 \\
$^{177}$Pt & A. Siivola & Nucl. Phys. &\cite{1966Sii01}& FE & Berkeley & USA &1966 \\
$^{178}$Pt & A. Siivola & Nucl. Phys. &\cite{1966Sii01}& FE & Berkeley & USA &1966 \\
$^{179}$Pt & A. Siivola & Nucl. Phys. &\cite{1966Sii01}& FE & Berkeley & USA &1966 \\
$^{180}$Pt & A. Siivola & Nucl. Phys. &\cite{1966Sii01}& FE & Berkeley & USA &1966 \\
$^{181}$Pt & A. Siivola & Nucl. Phys. &\cite{1966Sii01}& FE & Berkeley & USA &1966 \\
$^{182}$Pt & G. Graeffe & Ann. Acad. Sci. Fenn. &\cite{1963Gra01}& FE & Uppsala & Sweden &1963 \\
$^{183}$Pt & G. Graeffe & Ann. Acad. Sci. Fenn. &\cite{1963Gra01}& FE & Uppsala & Sweden &1963 \\
$^{184}$Pt & G. Graeffe & Ann. Acad. Sci. Fenn. &\cite{1963Gra01}& FE & Uppsala & Sweden &1963 \\
$^{185}$Pt & G. Albouy & J. Phys. Radium &\cite{1960Alb01}& LP & Orsay & France &1960 \\
$^{186}$Pt & V.I. Baranov & Bull. Acad. Sci. USSR &\cite{1961Bar01}& SP & Dubna & Russia &1961 \\
$^{187}$Pt & V.I. Baranov & Bull. Acad. Sci. USSR &\cite{1961Bar01}& SP & Dubna & Russia &1961 \\
$^{188}$Pt & R.A. Naumann & Phys. Rev. &\cite{1954Nau01}& LP & Harvard & USA &1954 \\
$^{189}$Pt & W.G. Smith & Phys. Rev. &\cite{1955Smi01}& LP & Berkeley & USA &1955 \\
$^{190}$Pt & H.E. Duckworth & Phys. Rev. &\cite{1949Duc01}& MS & Middletown & USA &1949 \\
$^{191}$Pt & G. Wilkinson & Phys. Rev. &\cite{1948Wil01}& LP & Berkeley & USA &1948 \\
$^{192}$Pt & A.J. Dempster & Nature &\cite{1935Dem01}& MS & Chicago & USA &1935 \\
$^{193}$Pt & G. Wilkinson & Phys. Rev. &\cite{1948Wil01}& LP & Berkeley & USA &1948 \\
$^{194}$Pt & B. Fuchs & Naturwiss. &\cite{1935Fuc01}& AS & Berlin & Germany &1935 \\
$^{195}$Pt & B. Fuchs & Naturwiss. &\cite{1935Fuc01}& AS & Berlin & Germany &1935 \\
$^{196}$Pt & B. Fuchs & Naturwiss. &\cite{1935Fuc01}& AS & Berlin & Germany &1935 \\
$^{197}$Pt & J.M. Cork & Phys. Rev. &\cite{1936Cor01}& LP & Berkeley & USA &1936 \\
$^{198}$Pt & A.J. Dempster & Nature &\cite{1935Dem01}& MS & Chicago & USA &1935 \\
$^{199}$Pt & E. McMillan & Phys. Rev. &\cite{1937McM01}& LP & Berkeley & USA &1937 \\
$^{200}$Pt & L.P. Roy & Phys. Rev. &\cite{1957Roy02}& NC & Chalk River & Canada &1957 \\
$^{201}$Pt & J. Facetti & Phys. Rev. &\cite{1962Fac01}& LP & Mayaguez & Puerto Rico &1962 \\
$^{202}$Pt & S. Shi & Z. Phys. A &\cite{1992Shi01}& LP & Shanghai & China &1992 \\
$^{203}$Pt & S.T. Steer & Phys. Rev. &\cite{2008Ste01}& PF & Darmstadt & Germany &2008 \\
$^{204}$Pt & S.T. Steer & Phys. Rev. &\cite{2008Ste01}& PF & Darmstadt & Germany &2008 \\
 &  &  & &  &  & & \\
&  &  & &  &  & & \\
 \\
\end{longtable}

\end{document}